\documentclass[a4paper,11pt]{article}
\usepackage{jcappub}

\newcommand{\gadget}{{\small GADGET}}
\newcommand{\mpgadget}{{\small MP-GADGET}}

\newcommand{\Lya}{Lyman-$\alpha$}

\newcommand{\hMpc}{\, h \mathrm{Mpc}^{-1}}

\title{An Emulator for the Lyman-$\alpha$ Forest}

\author[a,b,1]{Simeon Bird,\note{Corresponding author}}
\author[c]{Keir K. Rogers,}
\author[c, d]{Hiranya V. Peiris,}
\author[e, f]{\\ Licia Verde,}
\author[d]{Andreu Font-Ribera}
\author[d]{and Andrew Pontzen}
\affiliation[a]{Department of Physics \& Astronomy, University of California, Riverside,\\900 University Avenue, Riverside, CA 92521, USA}
\affiliation[b]{Department of Physics \& Astronomy, Johns Hopkins University, 3400 N.\ Charles Street, Baltimore, MD 21218, USA}
\affiliation[c]{Oskar Klein Centre for Cosmoparticle Physics, Stockholm University,\\AlbaNova, Stockholm SE-106 91, Sweden}
\affiliation[d]{Department of Physics \& Astronomy, University College London,\\Gower Street, London WC1E 6BT, UK}
\affiliation[e]{Institut de Ci\`encies del Cosmos, University of Barcelona,\\ICCUB, Barcelona 08028, Spain}
\affiliation[f]{Instituci\'o Catalana de Recerca i Estudis Avan\c{c}ats,\\Passeig Llu\'is Companys 23, Barcelona 08010, Spain}

\emailAdd{sbird@ucr.edu}
\emailAdd{keir.rogers@fysik.su.se}
\emailAdd{h.peiris@ucl.ac.uk}
\emailAdd{liciaverde@icc.ub.edu}
\emailAdd{a.font@ucl.ac.uk}
\emailAdd{a.pontzen@ucl.ac.uk}

\abstract{
We present methods for interpolating between the 1-D flux power spectrum of the Lyman-$\alpha$ forest, as output by cosmological hydrodynamic simulations. Interpolation is necessary for cosmological parameter estimation due to the limited number of simulations possible. We construct an emulator for the \Lya~forest flux power spectrum from $21$ small simulations using Latin hypercube sampling and Gaussian process interpolation. We show that this emulator has a typical accuracy of $1.5\%$ and a worst-case accuracy of $4\%$, which compares well to the current statistical error of $3-5\%$ at $z < 3$ from BOSS DR9. We compare to the previous state of the art, quadratic polynomial interpolation.  The Latin hypercube samples the entire volume of parameter space, while quadratic polynomial emulation samples only lower-dimensional subspaces. The Gaussian process provides an estimate of the emulation error and we show using test simulations that this estimate is reasonable. We construct a likelihood function and use it to show that the posterior constraints generated using the emulator are unbiased. We show that our Gaussian process emulator has lower emulation error than quadratic polynomial interpolation and thus produces tighter posterior confidence intervals, which will be essential for future \Lya~surveys such as DESI.
}

\begin{document}

\maketitle

\section{Introduction}
Modern cosmological surveys have extracted a large fraction of the information
available from scales where structure formation is well described by perturbation theory \cite{Planck:2016}. Hence, to resolve
the unsolved problems of cosmology, such as the nature of the inflaton or the mass of the neutrino, we
turn to measurements of small scale structure. One such probe is the \Lya~forest flux power spectrum\footnote{Strictly speaking, the power spectrum of the transmitted flux.}, which
measures correlations between neutral hydrogen along sightlines to luminous quasars \cite{Croft:1997}.
As the hydrogen gas clusters within dark matter potential wells, these correlations also measure
dark matter structure formation and so contain information about the underlying cosmology \citep{McDonald:2004data}.
The \Lya~forest is a uniquely powerful probe of cosmology because it is able to constrain the matter power spectrum
over a wide range of scales, from those of the Baryon Acoustic Oscillations (BAO) to some of the smallest currently available, $k \sim 10$ Mpc$^{-1}$.
Furthermore, it measures regions with only moderate over-density of $1-100$, which are minimally affected by the uncertain astrophysics
of galaxy formation and thus can robustly infer primordial fluctuations \cite{Viel:2013}.

The \Lya~forest 1D flux power spectrum, which measures correlations along the line of sight to the quasar, has been used to measure cosmological parameters \citep{Seljak:2006, Viel:2006},
measure the primordial perturbations \citep{Bird:2011}, constrain the number of neutrino species and
the neutrino mass \citep{PD2016, Rossi:2015}, the temperature of dark matter \citep{Viel:2013}, the redshift of reionization \cite{Nasir:2016, Boera:2018} and the mass of fuzzy dark
matter \cite{Irsic:2017b, Armengaud:2017}. The current best measurements of cosmological parameters use large samples of
low resolution quasar spectra obtained by the Sloan Digital Sky Survey (SDSS), both SDSS-II \citep{McDonald:2004data} and SDSS-III \citep{PD2013, PD2015, Chabanier:2018}.
Yet tighter constraints can be achieved by adding information on the thermal history of the intergalactic medium (IGM) from higher resolution spectra \cite{Irsic:2017, Yeche:2017}. On larger scales ($0.01 - 0.1 \hMpc$), the position of the Baryon Acoustic Oscillation peak in the 3D \Lya~forest correlation function has been used to constrain the expansion rate at $z > 2$ \cite{Slosar:2011, Busca:2013, Bautista:2017}.

In this paper we are concerned not with inferring the flux power spectrum from a survey,
but in using current measurements to estimate cosmological parameters. To do this, it is necessary to model
the growth of cosmological structure, including the distribution of
the gas in the IGM, on scales smaller than those described by linear perturbation theory. Cosmological hydrodynamic simulations are
the only method able to model this process at the percent level accuracy required by current data.
Unfortunately, Lyman-$\alpha$ forest simulations, while much cheaper than simulations with a full galaxy formation model, are several orders of magnitude more computationally intensive than evaluating a cosmological perturbation theory\footnote{The small simulations used in this paper required $\sim 400$ core-hours per simulation on XSEDE's Stampede 2 machine. A resolved $2\times 1024^3$ simulation would take about $35000$ core-hours on the same machine.}. This sharply limits the number available,
conflicting with the $\sim 10^6$ cosmological likelihood samples required in a standard Markov chain Monte Carlo (MCMC) approach to estimate parameter values.
We thus wish to find a way to choose a small number of points at which we can sample parameter space,
evaluate the model at these values, and interpolate between them with maximum accuracy for a given number of simulations.

Similar problems exist in other branches of science. In particular, engineers are frequently confronted with the need
for finding the parameter values which optimize the results of a complex non-linear model for, e.g.~the flow
of air over an aircraft wing. Techniques for interpolating between sparse model outputs are known
there as \textit{surrogate modelling}, with the interpolating routine itself known as a \textit{surrogate} \citep{Sacks_1989}.
Reviews of these techniques as applied in engineering can be found, for example,
in refs.~\cite{Queipo_2005, forrester2008engineering, forrester2009recent}.

Engineers, who for reasons of economy must ultimately construct only one type of aircraft, are concerned with finding a point estimate of a single optimal design. Cosmologists are by contrast uninterested in a point estimate, wishing instead to construct accurate confidence intervals on cosmological parameters.
This means that accurately quantified errors in the emulator should produce only loss of precision rather than a biased measurement.
However, in order to extract near-optimal constraints, an emulator must be accurate in a wider region of parameter space
corresponding to the measurement uncertainty.

Previous \Lya~forest analyses \citep{Viel:2005, Bird:2011, PD2013, PD2015} have interpolated using a simpler scheme:
quadratic polynomial interpolation in each parameter, around a central ``best guess''
simulation.\footnote{Although see ref.~\cite{McDonald:2005pk} for a different approach.}
While sufficiently accurate for current data, this scheme has several limitations, which motivate
the improvements presented in this paper.
Most importantly, polynomial interpolation provides no way to compute the uncertainty associated with the interpolation, except by running multiple costly test
simulations and assigning a global worst-case error. With the statistical error that will be achieved by future \Lya~forest surveys such as the Dark Energy Spectroscopic Instrument (DESI) \cite{FontRibera:2014}, the interpolation error must be quantified, controlled and accounted for precisely in order to minimize parameter bias and not significantly increase the resulting confidence intervals.

For a $P$ dimensional parameter space, interpolating each parameter separately keeps the number of simulations
linear in $P$. If each parameter requires $k$ simulations to estimate the coefficients of the polynomial ($k=4$ is a
reasonable choice for quadratic interpolation), the number of simulations required is $P\times k$. However,
because each parameter is interpolated independently, this procedure neglects correlations between
different parameters, which may be significant. These correlations could be captured using a
finely sampled regular grid in parameter space, but the $k^P$ simulations required would be prohibitive.

Similar interpolation problems have been encountered in cosmological surveys using other measurements of small-scale power.
This has led to the construction of \textit{cosmological emulators}, which use simulations to
make estimates which are uniformly accurate over a broad range in parameter space, and aim to be applicable
to a range of different experiments. The first application of emulation techniques to cosmology was ref.~\cite{Heitmann:2009},
who calibrated an emulator for the matter power spectrum with an accuracy of $1\%$. Refs.~\cite{Liu:2015, Petri:2015} calibrated
emulators for weak lensing observables, while ref.~\cite{Kwan:2013} calibrated an emulator for a galaxy halo occupation model. Emulation techniques
have also been used for the halo mass function \cite{McClintock:2018}, the galaxy correlation function \cite{Zhai:2018}, and the
21 cm power spectrum \cite{Jennings:2018}. Most similarly to our own work, ref.~\cite{Walther:2018}, who emulated the 1D flux power spectrum for
varying IGM thermal parameters at a fixed cosmology, to measure the temperature-density relation of the IGM. They did not make
use of the error estimate of the Gaussian process, instead calibrating error using test simulations. Ref.~\cite{Murgia:2018} used Gaussian
processes on top of a regular grid of simulations to place constraints on the dark matter temperature.

Here, we shall adapt several ideas from these earlier emulators to the problem of interpolating
the 1D \Lya~forest flux power spectrum for a full set of parameters, both cosmological and related to the astrophysics of the IGM.
We shall use Gaussian processes \citep{gpml}, a non-parametric Bayesian regression method,
to interpolate between simulations. Gaussian processes produce not just a point estimate of the interpolated function
at the desired parameter values, but an estimate of the interpolation error.
We shall add this error in quadrature to the data covariance matrix, allowing the likelihood to account for the uncertainty in our emulated values.
Instead of varying one parameter at a time, the parameter values sampled by our simulations will be chosen to fill a Latin hypercube in parameter space, which ensures that
each parameter value is unique within the simulation suite. This minimizes redundancy and maximizes the parameter support. Since our aim in this paper is to prove
our emulation methods, we use small simulations which are not fully numerically converged. This allows us to quickly iterate emulation tests without using
an undue amount of computer time. The lack of numerical convergence does not affect our results since we only compare our emulator results to test simulations
with the same resolution.

A major advantage of our techniques is that they are designed to allow later optimization of the emulator with the addition of further simulations.
Our algorithms and techniques for this Bayesian emulator optimization are presented in a companion paper, ref.~\cite{Rogers:2019}. Bayesian emulator optimization allows building
the interpolating function iteratively, focusing extra parameter evaluations in regions of high posterior probability and high interpolation uncertainty where they can be most useful.

In this work, we have constructed our emulator for the \Lya~forest 1D flux power spectrum. Several of the techniques developed here
are applicable to other datasets, such as weak lensing surveys, galaxy clustering and future 21cm measurements. These include especially the optimization techniques presented in our companion paper, our use of Latin hypercube sampling which maximizes the coverage of simulations in parameter space and our investigation of the conditions
under which an emulator produces unbiased results.

We present our methods in section \ref{sec:methods}. We describe the simulations used in section~\ref{sec:sims}, the parameters of the emulator in section~\ref{sec:simparams}, the construction of the emulator and the Gaussian process kernel in section~\ref{sec:gps} and the Latin hypercube sampling strategy in section \ref{sec:hyper}. Section~\ref{sec:likelihood} describes our likelihood function, including the Gaussian process error.

Section \ref{sec:results} presents our results. section \ref{sec:emuacc} shows the accuracy of our emulator by comparison to test simulations.
We compare our Gaussian process emulator explicitly to a quadratic polynomial emulator with the same number of points in section~\ref{sec:quadacc}.
Section \ref{sec:likeresults} shows posterior distributions given mock data. We conclude in section \ref{sec:conclusions} and in
appendix~\ref{sec:singleparam} we show the effect on the flux power spectrum of varying each emulator parameter.



\section{Methods}
\label{sec:methods}

In this section we describe our methods for building a \Lya~forest emulator. Section \ref{sec:sims} describes the simulations we perform. Section \ref{sec:simparams} describes the parameter space they sample, including our method for sampling the mean flux in post-processing. These sections describe standard methods for \Lya~forest simulations and so readers who are familiar with them may wish to skip to sections \ref{sec:gps} and \ref{sec:hyper} where we describe how we build our emulator from these simulations. Section \ref{sec:gps} describes how we interpolate between our simulations, while section \ref{sec:hyper} describes how we choose which simulation points to sample. We describe both our new Gaussian process based emulator and, for comparison, a standard quadratic polynomial emulator similar to that used by ref.~\cite{Bird:2011}.

\subsection{Simulations}
\label{sec:sims}

We use a number of cosmological hydrodynamic simulations performed with the massively parallel code \mpgadget \cite{yu_feng_2018_1451799}.\footnote{https://github.com/MP-Gadget/MP-Gadget/} \mpgadget~is a variant of the TreePM code \gadget-3, last described in \cite{Springel:2005}, which has been extensively modified to scale to $\sim 10^5$ MPI ranks and $25$ threads. Gravitational dynamics is followed using a Fourier transform based particle-mesh algorithm on large scales and a Barnes-Hut tree on small scales. Pressure forces from the gas are computed with the density-entropy formulation of smoothed particle hydrodynamics (SPH) with a cubic spline kernel and $33$ neighbours, following ref.~\cite{Springel:2005}. \mpgadget~also implements pressure-entropy SPH \cite{Hopkins:2013}, but we opt not to use it as it makes minimal difference for the low-density, largely adiabatic, gas of the \Lya~forest \cite{Bird:2014}. Following ref.~\cite{Viel:2004}, we use the simplified \texttt{quick\_lya} star formation criterion, which immediately turns all gas with an over-density $\Delta > 1000$ and a temperature $T < 10^5$ K into stars. Denser gas has a small volume filling fraction and thus contributes minimally to the \Lya~forest. The heating and cooling of the gas is computed following ref.~\cite{Katz:1996}, including an externally specified uniform meta-galactic ultra-violet background from ref.~\cite{Puchwein:2018}.

The purpose of this paper is to investigate methods for emulation, rather than produce a fully realized suite of simulations of the IGM. In order to allow us to quickly construct test emulators we have used relatively small, fast, simulations, containing $2\times 256^3$ dark matter and SPH particles within a $40$ Mpc/h (comoving) box. This small box size barely covers the scales probed by the BOSS flux power spectrum, and the low resolution means that small scales and high redshifts are not numerically converged \cite{Rossi:2015}. However, since in this paper we shall only compare our results to simulations of the same box size and resolution with different cosmological parameters, this will not affect our conclusions.

Our simulations are initialized with a Gaussian random field using the Zel'dovich approximation \cite{Zeldovich:1970}. Simulations are initialized at $z=99$ from linear matter power spectra generated using {\small CLASS} \cite{Lesgourgues:2011}. We use the same transfer function for both dark matter and baryon particles. Each simulation uses the same random realisation of cosmic structure, which has random phases, but individual modes are not scattered \cite{Anderson:2018}. Radiation density, assuming massless neutrinos, is included in the background evolution.

Each simulation generates $10$ output snapshots evenly spaced every $\Delta z = 0.2$ between $z=4.2$ and $z=2.2$, matching the redshift bins of BOSS DR9. These snapshots are post-processed by our artificial spectral generation code ``\texttt{fake\_spectra}''\footnote{\url{https://github.com/sbird/fake_spectra/}} \cite{FSFE:2017} to generate $32000$ \Lya~absorption spectra parallel to the $x$-axis of the simulation box. The positions of the spectra are chosen randomly but designed such that they are the same for each simulation and snapshot. We finally Fourier transform each spectrum along the quasar sightline and average the flux power spectrum over all sightlines to compute the mean one-dimensional flux power spectrum of the \Lya~forest. We generate our flux power spectra with a pixel width of $10$ km/s (and an infinite resolution spectrograph), substantially finer than the $69$ km/s pixel width of the BOSS spectrograph, which means the pixel window function correction is negligible.

There is a subtlety regarding the binning of the flux power spectrum. The conventional units for the flux power spectrum are velocity units, physical km/s, which are related to the simulation box units of comoving Mpc$/h$ by the Hubble rate:
\begin{equation}
 1 \mathrm{km/s} = \frac{100 \sqrt{\Omega_\mathrm{m} (1+z)^3 + \Omega_\Lambda}}{ (1 + z)} \mathrm{Mpc}/h\,.
 \label{eq:kms}
\end{equation}
This relation depends on the matter density $\Omega_\mathrm{m}$, one of the parameters which we wish to include in our model. Thus, as we change $\Omega_\mathrm{m}$, the
specific modes sampled by each bin in the flux power spectrum also change. On scales close to the box size the low number of modes per bin induces some scatter in the flux power spectrum. In order to avoid this scatter affecting our emulator we emulate flux power spectra using bins spaced as a fraction of the simulation box. Once an emulated flux power spectrum has been generated we convert to velocity (km/s) units and interpolate onto the bins used by the BOSS 1D flux power measurement.

\subsection{Simulation parameters}
\label{sec:simparams}

In this section we describe the parameter space spanned by our simulations. \Lya~forest simulations include both cosmological parameters and parameters to model the uncertain thermal state of the gas, which may be affected by helium reionization.  Our main purpose in this paper is to test and improve the interpolation method used. Our choice of parameters is thus similar to earlier work, in particular ref.~\cite{PD2013}.
In total we have five simulation parameters for our emulator, three cosmological and two astrophysical. The cosmological parameters are defined in section~\ref{sec:cosmoparams} and are $n_s$, $A_\mathrm{P}$ and $h$. The astrophysical parameters are defined in section~\ref{sec:astroparams} and are $H_A$ and $H_S$. There are also two parameters for the mean flux which do not require simulations, $\tau_0$ and $d\tau_0$, discussed in section~\ref{sec:meanflux}. These parameters and their prior limits are summarized in Table~\ref{tab:parameters}.

\begin{table}
\begin{center}
\begin{tabular}{|l|c|c|}
\hline
Parameter & Lower Limit &  Upper Limit \\
\hline
$n_s$ & $0.8$ & $0.995$ \\
$A_\mathrm{P}$ & $1.2\times  10^{-9}$ & $2.6 \times 10^{-9}$ \\
$h$ & $0.65$ & $0.75$ \\
$H_A$ & $0.4$ & $1.4$ \\
$H_S$ & $-0.7$ & $0.1$ \\
$\tau_0 $ & $0.75$ & $1.25$ \\
$d \tau_0 $ & $-0.25$ & $0.25$ \\
\hline
\end{tabular}
\end{center}
\caption{Summary of emulator parameters and their prior limits.}
\label{tab:parameters}
\end{table}

\subsubsection{Cosmological parameters}
\label{sec:cosmoparams}

We model the primordial power spectrum with two cosmological parameters, a slope $n_s$ and an amplitude $A_\mathrm{P}$. In order to minimize degeneracies between the cosmological parameters, we have introduced the parameter $A_\mathrm{P}$, which describes the power spectrum amplitude on $8$ Mpc scales.
Our primordial power spectrum is thus given by
\begin{equation}
  P(k) = A_\mathrm{P} \left(\frac{k}{k_P}\right)^{n_s -1}\,,
\end{equation}
where we use a pivot of $k_P=2 \pi / 8 = 0.78$ Mpc$^{-1}$. We have chosen $8$ Mpc because it is close to the mid-point of the scales measured by the 1D BOSS \Lya power spectrum. These scales are $1.1\times 10^{-3} - 2.0 \times 10^{-2}$ s/km, corresponding to $k = 0.075 - 1.4$ Mpc$^{-1}$ at $z=2.2$ and $k = 0.094 - 1.7$ Mpc$^{-1}$ at $z=4.2$. $A_\mathrm{P}$ is related to the standard CMB power spectrum amplitude parameter, $A^\mathrm{CMB}_s$, which is evaluated at a pivot scale of $0.05$ Mpc$^{-1}$, by
\begin{align}
 A^\mathrm{CMB}_s &= \left(\frac{0.4}{2 \pi}\right)^{n_s-1} A_\mathrm{P} \,.
\end{align}
We also add one parameter for the expansion history. We choose to vary the dimensionless Hubble parameter $h$ while fixing $\Omega_\mathrm{m} h^2$, which is well-measured by the CMB at $z=1100$ \cite{Planck:2016} and by \Lya~BAO at $z = 2.3$ \cite{Bourboux:2017}. This means that changing $h$ also changes the total matter density $\Omega_\mathrm{m}$ and thus affects the mapping between km/s and Mpc$/h$ units, as described in section \ref{sec:sims}. We fix $\Omega_b = 0.0483$, as this is degenerate with the mean flux (see section~\ref{sec:meanflux}). We vary $n_s$ between $0.8$ and $0.995$, $A_\mathrm{P}$ between $1.2 - 2.6 \times 10^{-9}$ and $h$ between $0.65$ and $0.75$. These limits are chosen so that they are centered on the current cosmological measurements from the BOSS DR9 1D \Lya~forest;  $n_s = 0.929 \pm 0.012$, $\sigma_8 = 0.84 \pm 0.03$ (corresponding to $A_\mathrm{P} = 1.9 \times 10^{-9}$) \cite{PD2015}. The prior limits on $n_s$ and $A_\mathrm{P}$ are chosen so that our emulator encompasses a parameter space approximately double the three-$\sigma$ limits from the \Lya~forest in each parameter as inferred by ref.~\cite{PD2015}.


There is nothing in our modelling which requires our particular choice of cosmological parameters. As long as parameters affect the flux power spectrum in a continuous fashion and are not highly degenerate, interpolation should work as for our tests. For example, we may in future consider a second-order scale dependence or ``running'' to the primordial power spectrum, or non-cold dark matter models \cite{Viel:2013}.


\subsubsection{Astrophysical parameters}
\label{sec:astroparams}

We include several astrophysical parameters to model the thermal history of the IGM. Following ref.~\cite{Viel:2004, Bolton:2008}, we attempt to account for the uncertain effect of helium reionization on the IGM temperature by rescaling the photo-heating by a density-dependent factor
\begin{equation}
  \tilde{\epsilon} = H_A \epsilon \Delta^{H_S}\,,
\end{equation}
where $\Delta$ is over-density. $H_A$ and $H_S$ are free parameters of the emulator.\footnote{In ref.~\cite{Bolton:2008},  $\alpha$ and $\beta$ respectively.}
$H_A$ controls the IGM temperature at mean density, $T_0$. $H_S$ controls the
slope of the IGM temperature-density relation, $\gamma$. We choose to use these photo-heating rates, $H_S$ and $H_A$, rather than the output thermal state of the IGM, $\gamma$ and $T_0$. Our choice means that our parameter choices are all input parameters to the simulation code, which is essential for the refinement algorithm presented in ref.~\cite{Rogers:2019}. It also has strong practical advantages when implementing the emulator. However, our final parameter constraints will be in terms of $H_S$ and $H_A$ and must be related to constraints on $\gamma$ and $T_0$.\footnote{The alternative, using $\gamma$ and $T_0$ as emulator parameters, was implemented by ref.~\cite{Walther:2018}.} We fix the redshift of hydrogen reionization.

A change of $0.4$ in $H_A$ corresponds to a $3000$K shift in $T_0$, while a change of $0.2$ in $H_S$ corresponds to a $0.8$ shift in $\gamma$. We vary $H_S$ between $-0.7$ ($\gamma = 1.22$ at $z=2.2$) and $0.1$ ($\gamma = 1.6$ at $z=2.2$), which corresponds to the approximate theoretical range expected from helium reionization \cite{McQuinn:2009}. $H_A$ varies between $0.4$ ($T_0 = 7600$ K at $z=2.2$) and $1.4$ ($T_0 = 17000$ K at $z=2.2$).

\subsubsection{Mean flux}
\label{sec:meanflux}

In addition to the parameters of the simulation, we also marginalize over uncertainty in the observed mean flux in post-processing. Assuming photo-ionization equilibrium, the mean flux of the \Lya~forest is proportional to the ionization fraction of neutral hydrogen and thus degenerate with the uncertain amplitude of the meta-galactic ultra-violet background (UVB). Mean flux rescaling is implemented by multiplying the optical depth in each spectral pixel by a constant so that the mean flux matches the desired value
\begin{equation}
 \bar{\mathcal{F}} = \left<\exp\left(-A \tau_i\right)\right>\,,
\end{equation}
We generate ten sets of spectra for each simulation with different mean flux values, evenly spaced between the $68\%$ confidence intervals from ref.~\cite{Kim:2007} for each redshift bin, which are given by
\begin{equation}
 \bar{\tau} = (23 \pm 7) \times 10^{-3} (1+z)^{3.65 \pm 0.21}\,.
\end{equation}
The \Lya~forest 1D flux power spectrum measures the mean flux substantially more accurately than the results of ref.~\cite{Kim:2007}, so this represents a weak prior. We have explicitly checked that our results do not change if we decrease the number of mean flux samples. Our emulator operates on $10 N_\mathrm{sim}$ flux power spectrum realizations in total. We found the emulator accuracy did not improve by increasing the number of samples.\footnote{To avoid potential aliasing in the emulator, the mean flux values sampled are offset by a small uniformly distributed random number which is different for each simulation. In practice, this reduced the standard deviation of the Gaussian process by only $\sim 0.1$\%.}

The mean optical depth is observed to be a power law with redshift. Our likelihood function thus assumes a power law mean optical depth and marginalizes over the amplitude and slope of this power law. The mean flux at each redshift is thus given by
\begin{equation}
\tau_i = \tau_0 (1+z)^{d\tau_0} \times 23 \times 10^{-3} (1+z)^{3.65}\,,
\end{equation}
where $\tau_0$ and $d\tau_0$ are the amplitude and slope of the redshift evolution of the mean flux. They have the parameter limits $0.75$ to $1.25$ and $-0.25$ to $0.25$, respectively. Given values for $\tau_0$ and $d\tau_0$, the mean flux in each redshift bin, $\tau_i$, is computed and an emulated flux power spectrum generated in each redshift bin. The same emulator is thus able to use more general parameterisations of the evolution of the mean flux with redshift, including marginalizing over a separate mean flux in each redshift bin.

\subsection{Interpolation and construction of the emulator}
\label{sec:gps}

The central purpose of this paper is to introduce a new Gaussian process based emulator for the \Lya~forest.
Gaussian processes can be viewed as a framework for Bayesian function prediction:
given a set of function evaluations $f_i$ at points $x_i$, the Gaussian process provides
a prior for estimating the function probability distribution at other points $x_j$: this
is the conditional probability $P(f(x_j) = f_j | f_i = f(x_i))$. The Gaussian process
prior itself is not particularly informative, corresponding to a constant function with
zero mean and unit variance. In order to use Gaussian processes in practice, we specify a covariance function.
This provides a probability structure for the function space, and describes correlations
between neighbouring points. For full generality, the covariance function may be learned
from the data itself \citep[e.g.~][]{Garnett:2017}. Any other choice represents an
additional (but very weak) prior on the structure of the data and should be motivated by a physical
understanding of the problem in question.

In practice it is best to experiment with different covariance functions and see which most accurately models the data.
If one were fitting to noisy data, one would need to guard against over-fitting. However, since
in this case we are performing interpolation there is no noise to fit and thus no danger of over-fit.
We are free to choose the best covariance function we can find.

Covariance functions generally have one or more free (hyper-) parameters.
These are estimated using optimization to maximize the marginal log-likelihood of the data given a particular hyperparameter. For further details on Gaussian process hyperparameter optimization methods, see ref.~\cite{gpml}, Eq. 5.4.

After some experimentation, we chose a radial basis function (RBF) kernel, combined with a dot, a.k.a. linear kernel.
The radial basis function (RBF) kernel\footnote{Sometimes referred to, somewhat confusingly, as the Gaussian process or squared exponential kernel.}
assumes that the probability that a function has a given value is given by a normal distribution
centered on the value at the nearest points, with the length-scale of the variance a free parameter
\begin{equation}
  K(x_i,x_j) = \exp\left(-\frac{|x_i - x_j|^2}{2l^2}\right)\,.
  \label{eq:rbfkernel}
\end{equation}
This kernel is a good default; it prefers smooth functions, like those which often occur in physical problems, yet it is flexible enough to accommodate many functional forms. The kernel weight is peaked locally, which means it avoids imposing strong priors on the global shape of the function.

We augment the RBF kernel with a dot kernel. This kernel is equivalent to Bayesian linear
interpolation. We include it because linear interpolation has already been shown to perform well
for the 1D \Lya~forest flux power spectrum \cite{Irsic:2017}. Since our Gaussian process emulator includes linear interpolation as a special case, it should perform at least as well.
Our final kernel is thus
\begin{equation}
  K(x_i,x_j) = \sigma_0^2 \exp\left(-\frac{|x_i - x_j|^2}{2 l^2}\right) + \sigma_i^2 |x_i.x_j|\,,
\end{equation}
where $l$, $\sigma_0$ and $\sigma_i$ are free hyper-parameters of the kernel, found by optimizing for the training data. In our Gaussian process emulator, the optimal values of these hyper-parameters were $l = 2.5 - 3$, $\sigma_0^2 = 0.85 - 1.2$ and $\sigma^2_i = 0.3 - 0.04$, with different values within each range depending on redshift. We found that in practice interpolation accuracy is insensitive to small changes in the hyper-parameters. Our interpolation code, flux power spectra, and simulation parameters are publicly available at \url{https://github.com/sbird/lya_emulator/} and \url{https://github.com/sbird/lya_emulator_training/}.

We achieved similar results with other standard kernels, in particular the Mat\'{e}rn $3/2$ kernel \cite{Matern:1960, gpml}, the rational quadratic kernel, the exponential kernel and higher powers of the linear kernel, which are equivalent to higher order polynomial regressions \cite{gpml}. We found, however, that the increased ability of the emulator to fit functions was compensated by an increased difficulty in estimating the kernel hyper-parameters and so more complex kernels did not improve emulator performance. We thus chose to use the simplest kernel that adequately predicted test simulations. In principle we should also allow separate RBF length scale hyper-parameters for each emulator parameter. However, in our emulator the prior ranges are chosen so that the flux power spectrum changes by a similar magnitude across the range of each parameter (see Appendix~\ref{sec:singleparam}) and so one length-scale suffices.

Our problem differs from the treatment of ref.~\cite{gpml} in one important respect; we are attempting to emulate a multi-output vector. We could build a Gaussian process for each bin of the flux power spectrum. This would imply optimizing hyper-parameters separately at each value of $k_F$ and $z$. Both for reasons of computational efficiency and to ensure that the emulator hyper-parameters are better constrained, we instead optimize hyper-parameters separately in each redshift bin. Thus we have $11$ separate Gaussian processes in our emulator, corresponding to the BOSS DR9 redshift outputs at $z=2.2, 2.4,...4.0, 4.2$. We considered using a single Gaussian process for the entire emulator, but found that this was less accurate; the hyper-parameters evolve substantially with redshift.\footnote{Note this also differs from many matter power spectrum emulators, which use separate hyper-parameters in each scale bin \cite{McClintock:2018}.}

In the Gaussian process implementation we use, \texttt{GPy} \cite{gpy2014},
the optimization technique used is the
(constrained) L-BFGS\footnote{Limited-memory Broyden–Fletcher–Goldfarb–Shanno.} optimizer implemented in \texttt{scipy}. We have also verified that
an MCMC optimizer implemented using \texttt{emcee} gave similar results.\footnote{We further tried
the optimizer of \texttt{scikit-learn}'s Gaussian process module. This did not correctly converge to the
optimum parameters at time of writing (\texttt{scikit-learn} version $0.20.0$).}

We divide our flux power spectra by the sample median, $\tilde{P}_\mathrm{F}$, to approximate the Gaussian process prior
of zero mean before performing our fit. Thus we are interpolating
\begin{equation}
  P_\mathrm{F,red} = P_\mathrm{F}/ \tilde{P}_\mathrm{F} - 1\,.
\end{equation}

\subsection{Selection of simulation points}
\label{sec:hyper}

In this section we describe our algorithm for choosing the parameter vectors to simulate. We describe two approaches. In section \ref{sec:lathyper}, the Latin hypercube design we use for our Gaussian process emulator, and, in section \ref{sec:singlegrid}, the single-parameter grid we use for our traditional quadratic polynomial emulator.

\subsubsection{Latin hypercube}
\label{sec:lathyper}

\begin{figure}
\includegraphics[width=0.45\textwidth]{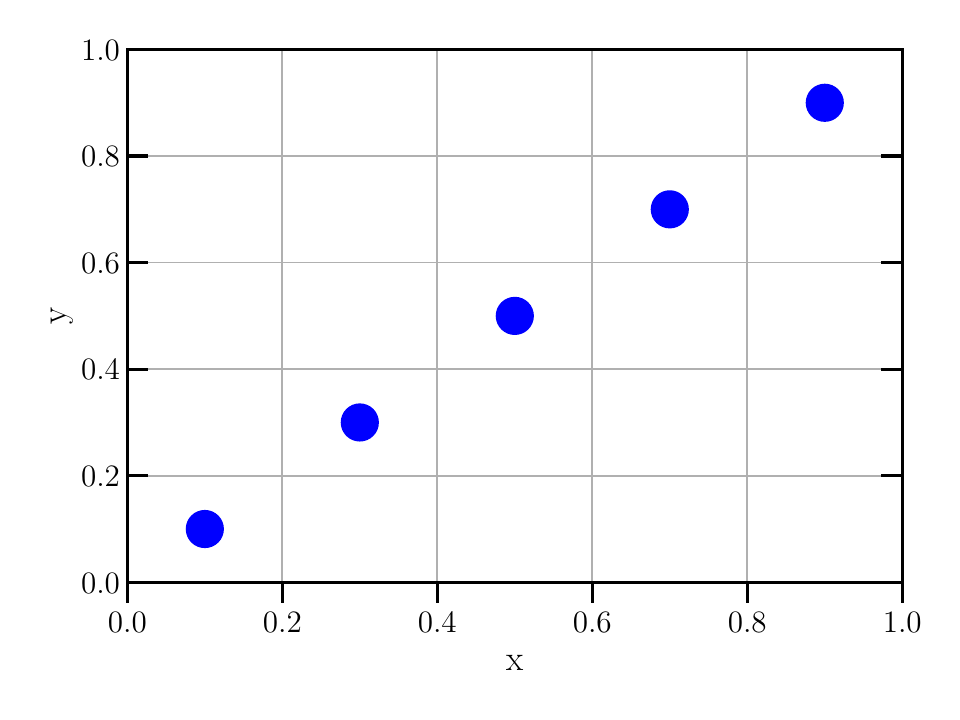}
\includegraphics[width=0.45\textwidth]{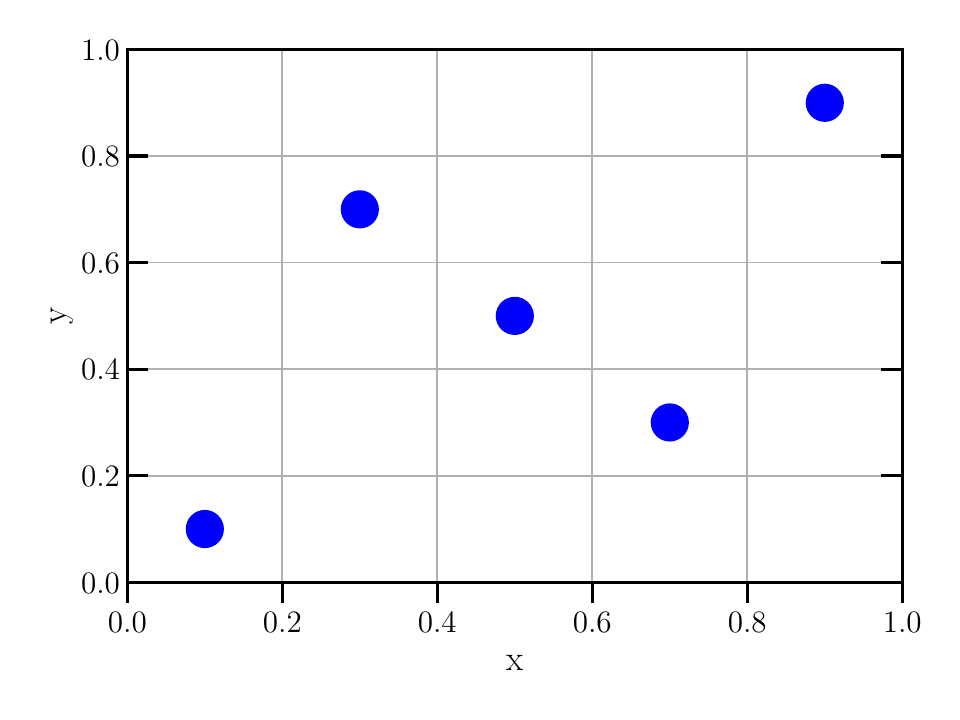}
  \caption{Two possible Latin hypercubes in a two dimensional parameter space normalized to the unit square.
  The left example covers only a one-dimensional subspace and hence would be rejected.
  The right example has support throughout the whole two dimensional parameter space and would produce a better emulator.}
  \label{fig:hypercube}
\end{figure}


The Gaussian process kernel described in section \ref{sec:gps} has an uncertainty roughly proportional to the distance between a parameter vector and the nearest simulation. In order to minimize the maximum error, we would like to spread our simulations evenly throughout parameter space. We would also like to avoid simulations clustered in low-dimensional sub-spaces, which would be unable to reproduce parameter variations elsewhere.

To ensure these properties, we choose the sampling points for our simulations using a Latin hypercube, as have been used extensively in cosmological emulators \cite{Heitmann:2009}. This structure divides a (normalized) parameter space into a regular grid, with as many rows in each dimension as sampling points. Thus for an emulator with $n$ dimensions containing $k$ simulations we would first create a $k^n$ grid. The samples are placed in turn such that no sample overlaps in any dimension. Each sample added thus blocks off one row in all $n$ dimensions.
This is equivalent to choosing a random permutation of the integers $0-k$ \cite{forrester2008engineering} and ensures that once the samples are projected to a single dimension, they evenly sample the unit interval. Note that there is no restriction on the number of simulations that can be placed within a given parameter space.


There are a large number of possible Latin hypercubes and not all of them ensure our desired properties. Figure~\ref{fig:hypercube} shows two, for $n=2$ and $k = 5$. We would like to avoid Latin hypercubes which fill only a one-dimensional subspace of parameter space, as in the left example. The most common technique for ensuring space-filling properties is to use orthogonal Latin hypercubes, which enforce uniform sampling not just in each parameter, but in each $2$ or $3$-dimensional subspace. However, these designs are inflexible and not suited to adding extra simulation points.

Instead, we use a Monte Carlo approach. We generate $10,000$ hypercubes at random and use the one which maximizes our space-filling metric, discarding the rest. Our space-filling metric is the sum of the Euclidean distances between each point in the hypercube and its closest neighbour. This approach is simple, computationally efficient, and allows for the addition of extra simulation points to the design in a later refinement step, as in our companion paper ref.~\cite{Rogers:2019}. The right panel of Figure \ref{fig:hypercube} shows a simple Latin hypercube generated using this procedure, demonstrating that these samples achieve good coverage of parameter space.


\subsubsection{Quadratic emulator: single parameter grid}
\label{sec:singlegrid}

\begin{figure*}
\centering
\includegraphics[width=0.45\textwidth]{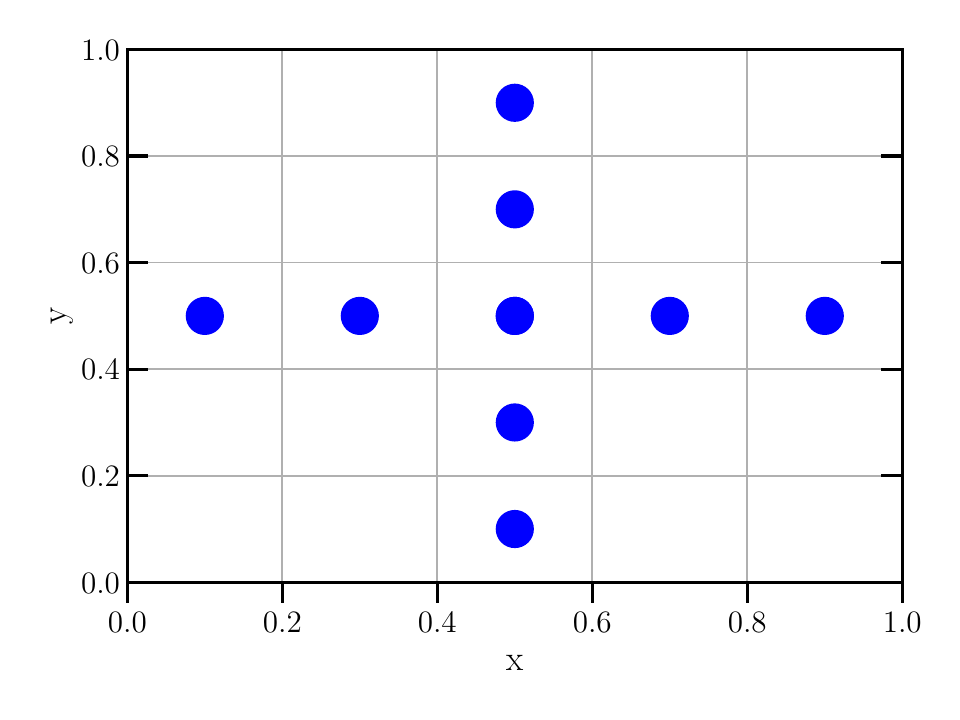}
  \caption{Locations of the sampling points for a quadratic emulator in a (normalized) two-dimensional parameter space for a ``best-fit'' simulation at $(0.5, 0.5)$. Our emulator uses $4 n +1 $ simulations.}
  \label{fig:quadratic}
\end{figure*}

In order to compare our new emulator techniques to prior work, we also generate a `quadratic' emulator similar to ref.~\cite{Bird:2011}. First a ``best-fit'' set of parameters is simulated, chosen to lie at the midpoint of each dimension in the $n$-dimensional parameter space prior volume. For our quadratic emulator, this parameter combination is: $\tau_0 = 1$, $d\tau_0 = 0$, $n_s = 0.8975$, $A_\mathrm{P} = 1.9 \times 10^{-9}$, $H_S = -0.3$, $H_A = 0.9$, $h = 0.7$. Each simulation varies only one parameter from this midpoint. Each bin of the resulting flux power spectra is fit with a quadratic polynomial, and interpolation evaluates each quadratic polynomial at the new parameters. We run four simulations for each parameter, for a total of $4 n + 1$ simulations. Although this procedure may seem simplistic, and does not capture variations involving multiple parameters, it can work reasonably well \cite{Bird:2011}, because the flux power spectrum varies almost linearly in the \Lya~forest parameters (see appendix~\ref{sec:singleparam}).

\subsection{Likelihood}
\label{sec:likelihood}

In this section we describe our likelihood function. The previous sections have described how we can generate a flux power spectrum for arbitrary input cosmological parameters. While we will see that our procedure can do this with a reasonable degree of accuracy, an important advantage of the Gaussian process is that it also provides an estimate of the error in the interpolation. This estimate can be used to propagate interpolation error into posterior probabilities. In principle (as long as the actual interpolation error is approximately Gaussian), this should cause moderate interpolation error to increase uncertainty, without leading to a bias in the derived cosmological parameters.  Our likelihood is sampled using the \texttt{emcee} package \cite{emcee}.

We use a simple quadratic (log) likelihood function, given by
\begin{align}
 \log \mathcal{L}(p_k) &= -\frac{1}{2} C^{-1}_{i j}(p_k) dP_i(p_k) dP_j(p_k) - \frac{1}{2}\log \mathrm{det}\,C_{i j} (p_k)  \\
 dP_i(p_k) &= P^\mathrm{sim}_F(k_i, p_k) - P_F^\mathrm{data}(k_i)\,,
 \label{eq:likelihood}
\end{align}
where $C_{i j} (p_k)$ is the total covariance matrix and $P^\mathrm{sim}_F(k_i, p_k)$ the simulated flux power spectrum for a set of parameters $p_k$. $P^\mathrm{data}(k_j)$ is the flux power spectrum of the mock data. We impose a hard prior on all parameters that they must be within the parameter bounds of the simulated emulator given in Table~\ref{tab:parameters}.

The covariance matrix $C_{i j}$ is the sum of the total BOSS DR9 covariance from ref.~\cite{PD2013} and the emulator error generated by the Gaussian process
\begin{equation}
 C_{i j}(p_k) = C_{i j}^\mathrm{BOSS} + C_{i j}^\mathrm{GP} (p_k)\,.
\end{equation}
While $C_{i j}^\mathrm{BOSS}$ does not depend on cosmology, $C_{i j}^\mathrm{GP} (p_k)$ does, depending primarily on the distance between $p_k$ and the nearest simulation point, and has an amplitude which is related to a hyper-parameter of the Gaussian process ($\sigma_0$). Because $C_{i j}(p_k)$ thus depends on $p_k$, every likelihood evaluation requires a matrix inversion. Furthermore, in principle $\log \mathrm{det}\,C_{i j}$ is not constant and cannot be absorbed into the likelihood normalization, although in practice we find that it does not vary significantly. The need to invert matrices slows down likelihood computation only moderately: evaluating our Gaussian Process likelihood is a factor of two slower than the quadratic likelihood.

For a Gaussian process with single-valued output, the covariance matrix would be diagonal and given by $\sigma^2(p_i)$, the Gaussian process expected error at parameters $p_i$. However, we found a substantial correlation in the actual error between each $k$-bin. This can be understood because, as detailed in section~\ref{sec:gps}, our Gaussian process uses the same hyper-parameters to estimate all scales in a single redshift bin. There are also substantial correlations between different $k$-bins in the 1D \Lya~flux power spectrum, because it probes non-linear scales. In order to ensure that we do not bias the likelihood, we conservatively assume that all $k$-bins at the same redshift are maximally correlated, minimizing the information content.
However, because each redshift bin has separate hyper-parameters, we approximate the correlations in emulator error as zero if two data points are in different redshift bins. The covariance matrix for the emulator is thus
\begin{equation}
 C_{i j} = \sigma(z_i, p_i) \delta_{z_i z_j}\,,
\end{equation}
where $\sigma(z_i, p_i)$ is the error reported by the Gaussian process at the parameter vector $p_i$ at redshift bin $z_i$, and $\delta_{z_i z_j}$ is unity if $i$ and $j$ are at the same redshift and zero otherwise.

As detailed in section~\ref{sec:sims}, we generate an interpolated flux power spectrum $P^\mathrm{sim}_F(k_i, p_k)$ at values of $k^\mathrm{box}_i$ which are multiples of the simulation box (``Fourier'' units). Thus, we rebin the flux power spectrum so that it is evaluated at values of $k_i$ which are the same as those in BOSS DR9. Since our simulation boxes are relatively small, the largest scales measured by BOSS are on scales larger than the simulation box and are omitted from the likelihood.

As our simulations are not resolved, and to preserve the option of blinding any later analysis with larger simulations, we do not use the actual BOSS flux power spectrum measurement for $P^\mathrm{data}(k_j)$. Instead we have performed a series of $6$ test simulations. The parameters of these test simulations were generated at random using a Latin hypercube (as in the main emulator, section~\ref{sec:lathyper}) with slightly restricted cosmological parameter limits: $n_s = 0.9 - 0.99$ and $A_\mathrm{P} = 1.4 \times 10^{-9} - 2.6 \times 10^{-9}$. This ensures that our tests cover our prior volume but that some of our likelihood posteriors are separated from the emulator bounds.

\section{Results}
\label{sec:results}

\begin{figure*}
   \includegraphics[width=0.45\textwidth]{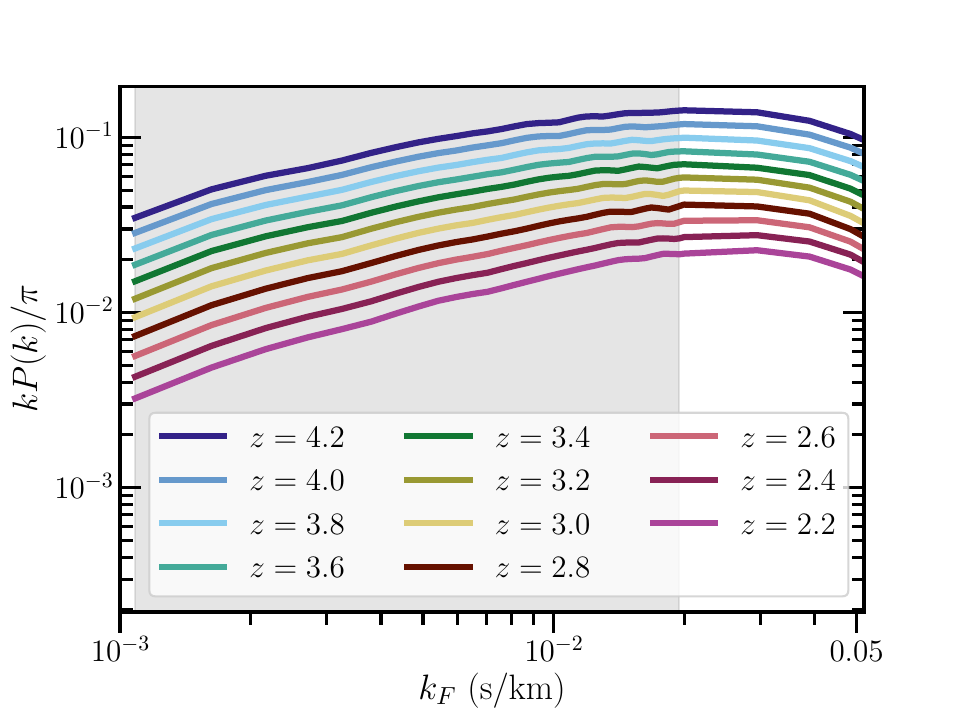}
   \includegraphics[width=0.45\textwidth]{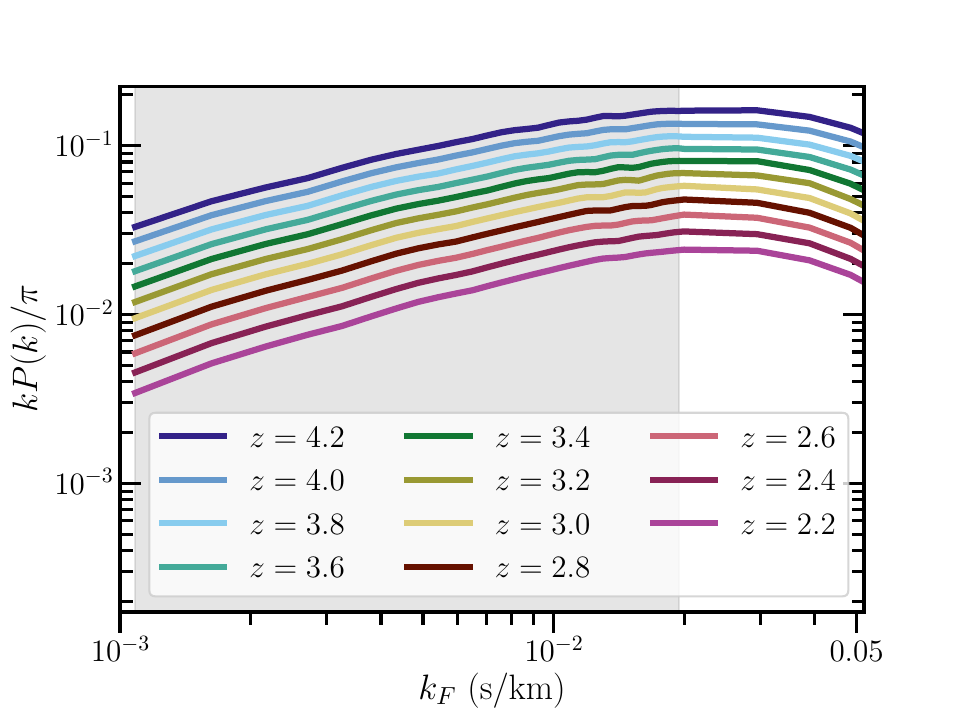}
  \caption{Flux power spectra extracted from two of our test simulations. Each line represents a different redshift bin. The grey shaded region shows the scales measured by BOSS.
  (Left) Test simulation parameters are $\tau_0 = 0.9$, $d\tau_0 = 0$, $n_s = 0.907$, $A_s=1.7\times 10^{-9}$, $H_A = 0.933$, $H_S = -0.633$, $h = 0.742$.
  (Right) Test simulation parameters are $\tau_0 = 0.9$, $d\tau_0 = 0$,  $n_s = 0.982$, $A_s=2.5 \times 10^{-9}$, $H_A = 0.667$, $H_S = -0.1$, $h = 0.658$.
  }
  \label{fig:testfpk}
\end{figure*}

In this section we describe the performance of our emulator methods. In section~\ref{sec:emuacc} we
describe the ability of our Gaussian process emulator to accurately interpolate the flux power spectrum. We compare to a quadratic polynomial emulator in section~\ref{sec:quadacc}. Then in section~\ref{sec:likeresults} we show posterior probabilities generated from likelihood functions using both the Gaussian process and quadratic emulators. We demonstrate that the true parameters lie within the $68\%$ confidence intervals of the likelihood when using our Gaussian process, with smaller parameter errors than the quadratic emulator. To aid the reader in visualising our results, figure~\ref{fig:testfpk} shows the flux power spectra from our test simulations, whose parameters provide the ground truth to which we compare our emulation and Markov Chain results.

\subsection{Gaussian process emulator accuracy}
\label{sec:emuacc}

\begin{figure*}
   \includegraphics[width=0.45\textwidth]{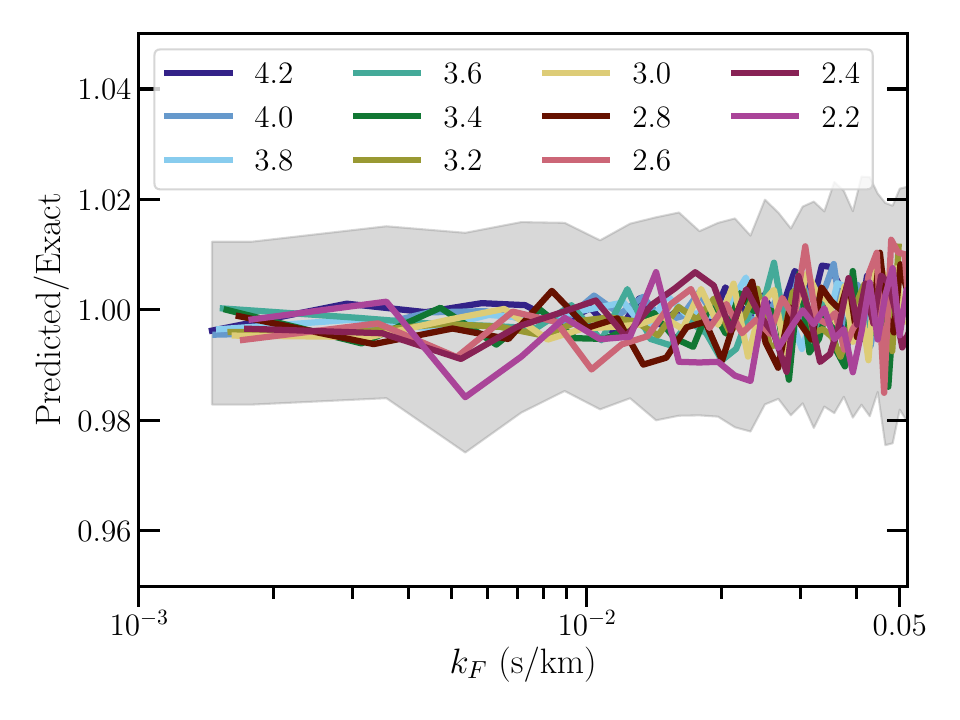}
   \includegraphics[width=0.45\textwidth]{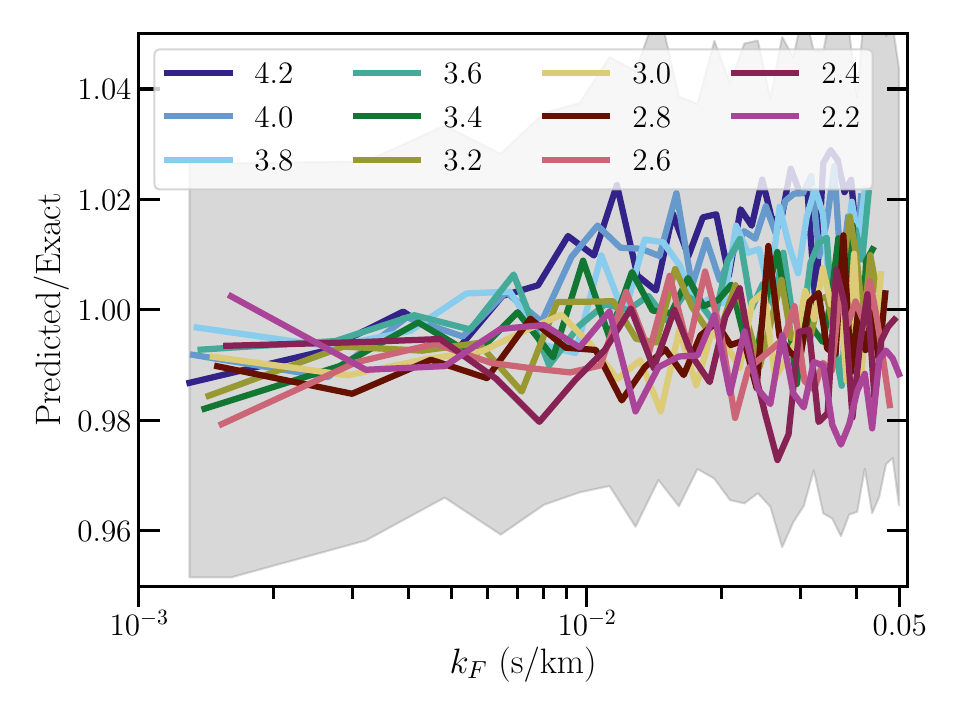}
  \caption{Accuracy of the Gaussian process emulator. We show the ratio between an emulated flux power spectrum and the true flux power spectrum from a simulation with the same set of cosmological parameters. Each line represents a different redshift bin. The grey band shows the expected error from the emulator. For clarity we show the maximum region covered by the $1-\sigma$ emulator error prediction. This is larger than the true difference, both because it combines multiple redshift bins and because the emulator slightly over-estimates the true error at these parameters.
  (Left) Test simulation parameters are $\tau_0 = 0.9$, $d\tau_0 = 0$, $n_s = 0.907$, $A_s=1.7\times 10^{-9}$, $H_A = 0.933$, $H_S = -0.633$, $h = 0.742$.
  (Right) Test simulation parameters are $\tau_0 = 0.9$, $d\tau_0 = 0$,  $n_s = 0.982$, $A_s=2.5 \times 10^{-9}$, $H_A = 0.667$, $H_S = -0.1$, $h = 0.658$.
  }
  \label{fig:nss8acc}
\end{figure*}

\begin{figure*}
   \includegraphics[width=0.45\textwidth]{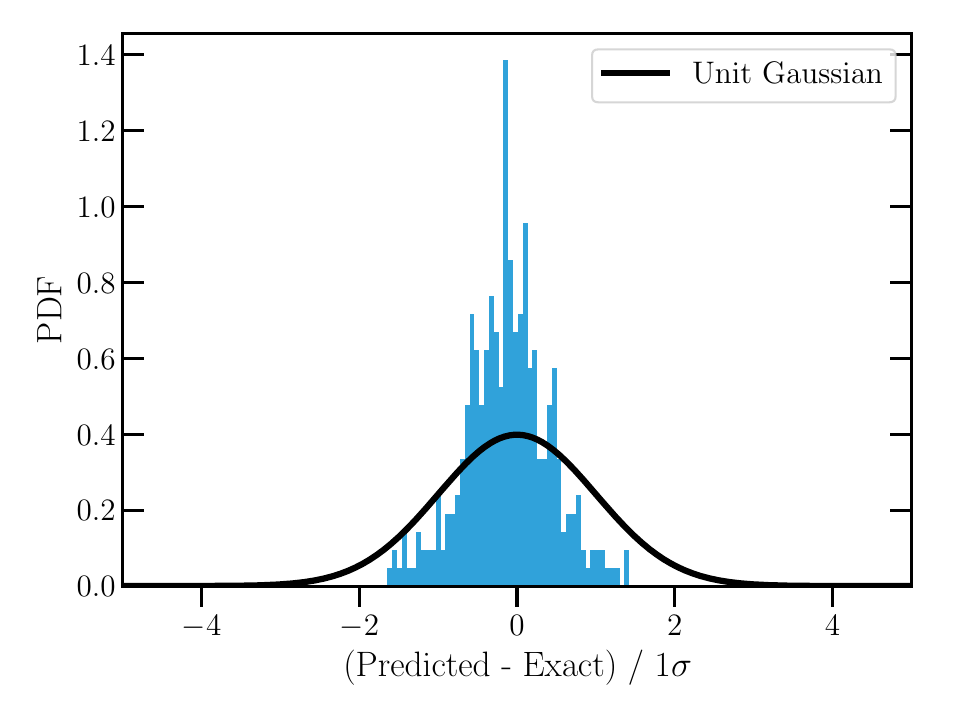}
   \includegraphics[width=0.45\textwidth]{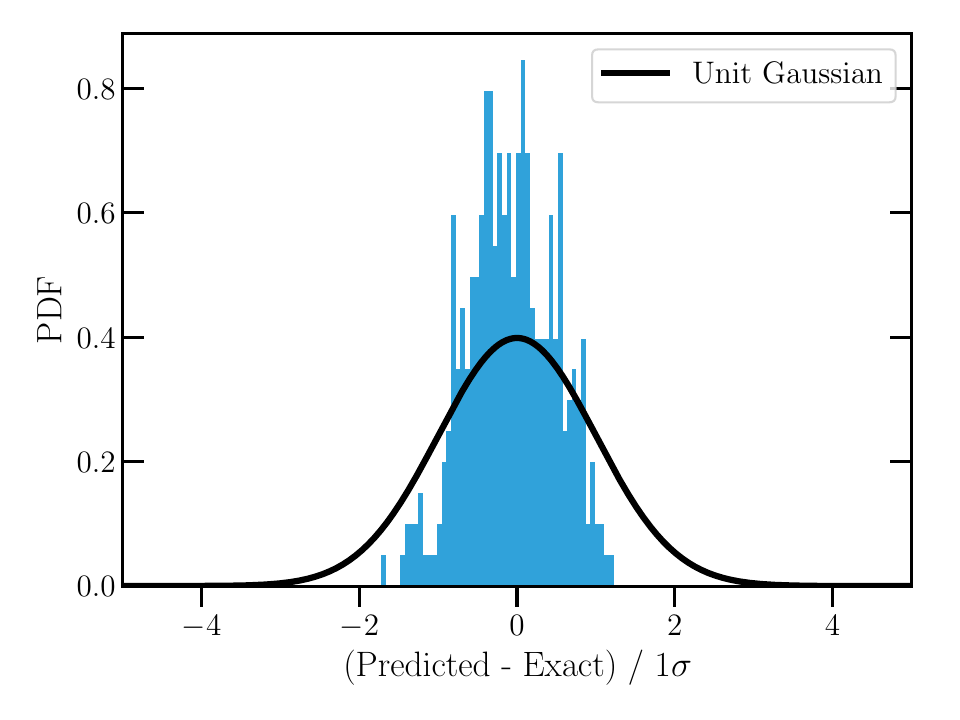}
  \caption{Distribution of the errors of the Gaussian process for the same test simulations in figure~\protect\ref{fig:nss8acc}. We show a histogram of the actual difference in each $k$ and $z$ bin between the flux power spectrum estimated from the emulator and the true flux power spectrum from a simulation, normalized by the $1-\sigma$ error expected from the emulator. Shown for comparison is a unit Gaussian. Thus an ideal error estimation would result in a match between the black line and the blue histogram. Each figure shows a different simulation with $11$ independent redshift bins and $31$ correlated $k$-bins. For these two simulations the Gaussian process is slightly over-estimating the true errors. In other test simulations the Gaussian process expected error is more accurate.
  (Left) Test simulation parameters are $\tau_0 = 0.9$, $d\tau_0 = 0$, $n_s = 0.907$, $A_s=1.7\times 10^{-9}$, $H_A = 0.933$, $H_S = -0.633$, $h = 0.742$.
  (Right) Test simulation parameters are $\tau_0 = 0.9$, $d\tau_0 = 0$,  $n_s = 0.982$, $A_s=2.5 \times 10^{-9}$, $H_A = 0.667$, $H_S = -0.1$, $h = 0.658$.
  }
  \label{fig:erracc}
\end{figure*}


Figure~\ref{fig:nss8acc}~shows the accuracy of a Gaussian process emulator using
the cosmological parameters for two test simulations. The parameter file for the
simulations used, as well as the flux power spectra, are available from
\url{https://github.com/sbird/lya_emulator_training/}. We tested the accuracy of
our Gaussian process emulator with six test simulations not included in the emulator,
as described in section~\ref{sec:likelihood}. From these six we show in the left panel
one simulation where the accuracy of the emulator was typical, and in the right panel
the test simulation where the accuracy was poorest. We show the differences in each
redshift bin between the flux power spectrum generated by the emulator and that
from a test simulation. The typical difference between the Gaussian process prediction and
the simulated flux power spectrum is of the order of $1\%$, with a worst-case
performance of $2\%$ (although the estimated emulator error,
shown by the grey band in figure~\ref{fig:nss8acc}, is larger). We further
tested the Gaussian process emulator using the input simulations, which were
reproduced to machine precision, and using the $21$ simulations run for the
quadratic emulator. The latter had similar performance to the other $6$
test simulations, and showed that the emulator is generally
accurate to $\sim 1\%$, with only $6 / 27$ simulations showing $>2\%$ inaccuracy.
For comparison, the current statistical error on the BOSS DR9 1D flux power spectrum
is minimal at around $\sim 3\%$ at $z=2.4$, rising to $6\%$ at $z=3.6$
and $\sim 10\%$ at $z \geq 4$. Figure~\ref{fig:nss8acc} thus demonstrates that
our emulator is capable of achieving the percent-level accuracy required for both current
and next-generation \Lya~surveys.

Notice that the residual errors are often correlated between neighbouring $k$-bins,
but that the largest and smallest scales are uncorrelated. Furthermore, there appears
to be little correlation between neighbouring redshift bins, justifying the assumptions
of our likelihood function. For this emulator, error is usually worst on small scales.
This has a simple explanation: small scales are more non-linear and thus inherently harder to model.
Furthermore, the IGM thermal history affects the shape of the flux power
spectrum only on small scales. Thus the emulator has to fit more complex variations
over more parameters at these scales. Recovery of the power spectrum appears to be worse
at high redshift in the right-hand plot of Figure~\ref{fig:nss8acc}, but this is a feature
specific to this test case: there is no general trend of interpolation error with redshift.

Figure~\ref{fig:erracc} shows histograms of the actual errors normalized by the expected error
from the Gaussian process. Also shown is a unit Gaussian. If the emulator's error estimation
was correct, the histograms should be close to the unit Gaussian. We can see that for these
two examples the Gaussian process is moderately over-estimating the true error. This is mostly
due to a statistical fluctuation in these two test simulations.\footnote{Note that the correlation between
neighbouring $k$-bins means that there are far fewer independent samples than bins.}
Other test simulations showed an error distribution closer to the unit Gaussian, and a histogram combining
errors from all our test simulations was much closer. None of our test simulations showed the
emulator under-estimating the error, but $3/27$ tests showed $> 1-\sigma$ offset between
the true error distribution and the unit Gaussian, indicative of a biased estimate of the
flux power spectrum. Overall, we found that it was more difficult for the Gaussian process
to accurately estimate errors than to produce a good estimate of the flux power spectrum.

\subsubsection{Emulator accuracy}

The overall accuracy of the emulator is controlled by the density of simulations. From the radial basis function kernel, eq.~\ref{eq:rbfkernel},
the emulator error scales like
\begin{equation}
 E  = E_0 \frac{\Delta x}{\sqrt{2} l}\,,
 \label{eq:eerror}
\end{equation}
where $\Delta x$ is the distance to the nearest simulation point, in units where all parameters are mapped onto the unit cube.
For our emulator the RBF length scale is $l = 2.5 - 3$, depending on the redshift. For evenly spaced simulations, the
maximum distance from any point to the nearest simulation is
\begin{equation}
 \Delta x \sim \sqrt{\frac{P}{N_\mathrm{sim}^2}}\,,
\end{equation}
where $N_\mathrm{sim}$ is the number of simulations and $P$ is the number of parameters. For us $N_\mathrm{sim} = 21$ and $P = 7$,
so $\Delta x \approx 0.13$. We achieve an average error of $1.5\%$, so if $l = 3$, $E_0 \approx 0.5$.
Inverting Eq.~\ref{eq:eerror} shows that the emulator error scales linearly with the number of simulations and that
a median error of $\sim 4\%$ would require only $8$ simulations. A median emulator error larger than the statistical error
in the data can present difficulties for parameter estimation, as the covariance matrix becomes highly parameter dependent,
in extreme cases leading to a multi-modal likelihood.

This calculation, however, neglects an important source of uncertainty in the emulator; the need to estimate the kernel hyper-parameters.
For sparse data it becomes extremely difficult to determine the optimal hyper-parameter and thus the emulator is inaccurate. Worse, because
the emulator error is itself a hyper-parameter this source of error is not included in the emulator's error estimate and thus
the posterior confidence intervals become biased. We have been unable to find firm guidance in the statistics literature as to what simulation
point density is necessary for a good hyper-parameter estimation.
A reasonable ansatz seems that the kernel length scale should be at least the average distance between simulations:
$\Delta x = \sqrt{P} / N = K l$. Note that $l$ corresponds to a physical scale of
variation in the flux power spectrum and so the number of simulations needed will scale with the prior parameter volume.
The value of $K$ is uncertain, and is likely to depend on the exact form of the kernel. As the hyper-parameters of our
Gaussian process emulator are well-determined, we know that $K \lesssim 20$. An earlier iteration of our test emulator
with a prior volume $5$ times larger did not have well-determined hyper-parameters, suggesting that for
our kernel $K \gtrsim 5$.

\subsection{Comparison to a quadratic emulator}
\label{sec:quadacc}

\begin{figure*}
    \includegraphics[width=0.45\textwidth]{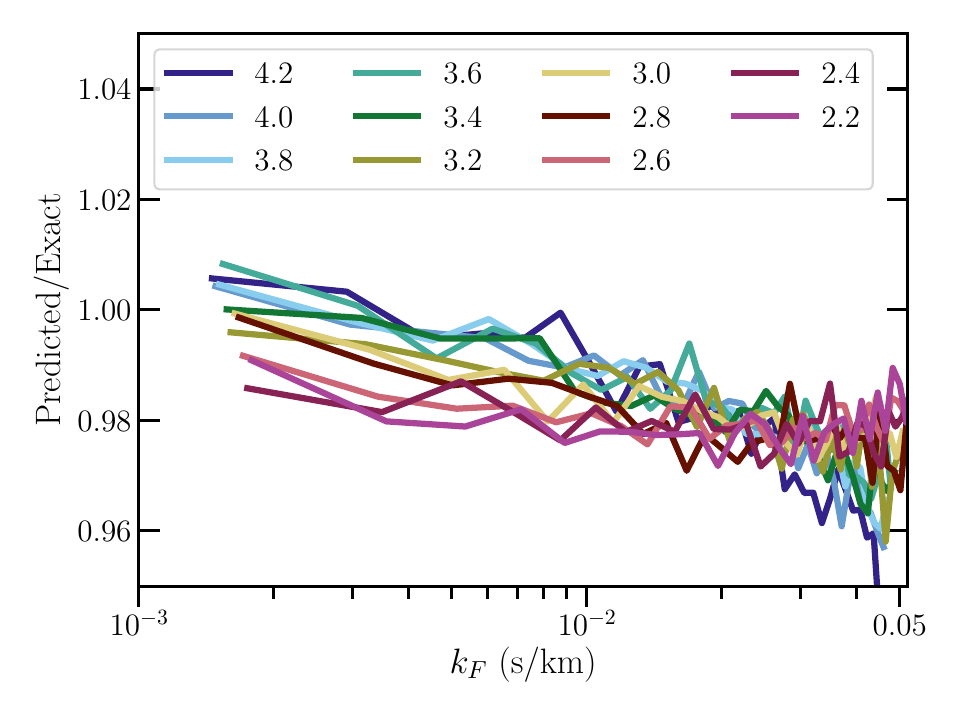}
    \includegraphics[width=0.45\textwidth]{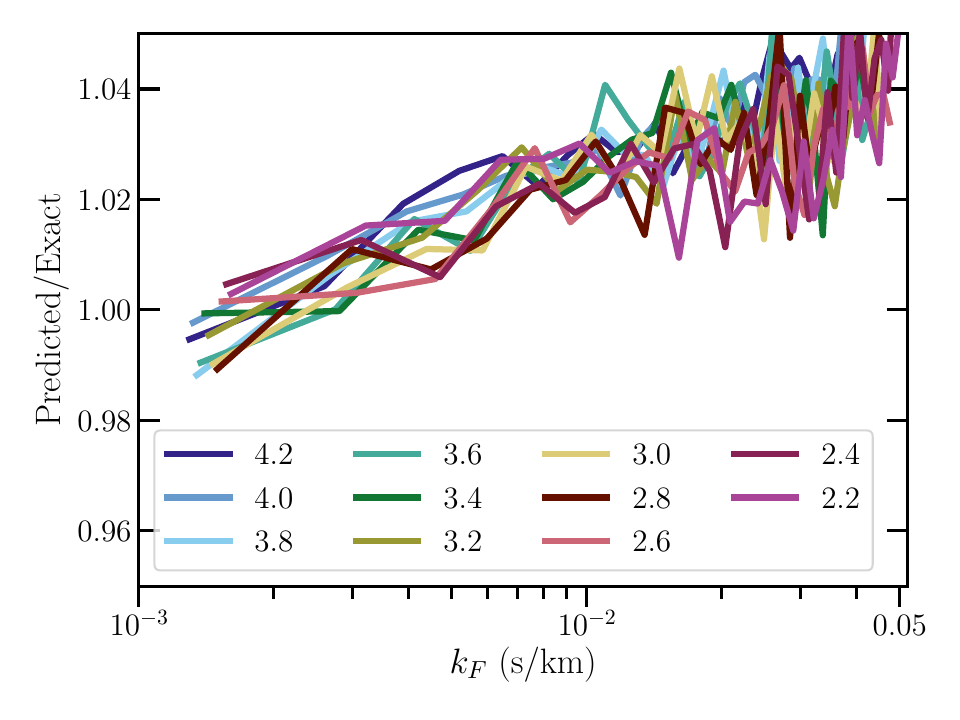}
  \caption{
  Accuracy of the emulator using quadratic polynomial interpolation. We show the ratio between an emulated flux power spectrum and the true flux power spectrum from a simulation with the same set of cosmological parameters. Each line represents a different redshift bin. Test simulation parameters are identical to those used in figure~\protect\ref{fig:nss8acc}.
  (Left) Test simulation parameters are $\tau_0 = 0.9$, $d\tau_0 = 0$, $n_s = 0.907$, $A_s=1.7\times 10^{-9}$, $H_A = 0.933$, $H_S = -0.633$, $h = 0.742$.
  (Right) Test simulation parameters are $\tau_0 = 0.9$, $d\tau_0 = 0$,  $n_s = 0.982$, $A_s=2.5 \times 10^{-9}$, $H_A = 0.667$, $H_S = -0.1$, $h = 0.658$.
  }
  \label{fig:quadraticacc}
\end{figure*}

Figure~\ref{fig:quadraticacc} shows the accuracy of a quadratic polynomial emulator, using
the same test simulations, prior parameter ranges and number of simulations as the Gaussian process shown
in figure~\ref{fig:nss8acc}. A comparison between figures~\ref{fig:nss8acc} and~\ref{fig:quadraticacc}
thus shows directly the relative performance of each emulator. We find that the overall accuracy
of our quadratic emulator at these test points is around $4\%$. Two of our six test simulations achieved $2\%$ accuracy.
This performance is clearly worse than the Gaussian process emulator. Furthermore,
figure~\ref{fig:quadraticacc} shows that the error is highly correlated between redshift bins, which
may present a difficulty for parameter estimation. Note that we have generated test spectra
at $\tau_0 = 0.9$, whereas the quadratic emulator is centered on $\tau_0 = 1$. The value of $\tau_0$
at which the quadratic emulator is evaluated strongly changes the overall accuracy. In fact, at $\tau_0 = 1$,
the quadratic emulator performs almost as well as the Gaussian process emulator. This is not surprising, as
the mean flux, $\tau_0$, is the single parameter which most affects the flux power spectrum. However, the
accuracy of the quadratic emulator degrades quickly as the mean flux changes.
As the quadratic emulator only varies one parameter at a time, all simulations except the central one
have their flux power spectra generated at $\tau_0 = 1$, and any correlation between power spectrum
shape variation from the mean flux and from cosmology is not captured. This sensitivity
to the mean flux is not seen in the Gaussian process emulator, which generates a series of flux power
spectra with different mean fluxes for every simulation.
Figure~\ref{fig:quadlikelihood} shows that posterior error on the mean flux is
around $\Delta \tau_0 = 0.1$ so the quadratic emulator is always inaccurate in some part of the $68\%$ confidence intervals.

We also note that the overall accuracy of the quadratic emulator is sensitive to the prior
parameter range. Although the quadratic emulator shown here has an accuracy of only $4\%$,
quadratic emulators used in prior work \cite[e.g~][]{Viel:2005, Bird:2011, PD2013} have achieved
better performance by using a restricted prior parameter range, or by carefully choosing
the simulated parameters to lie close to the best fit value of the model. However, this is not always possible, as
the emulator must encompass the full data posterior, and the best fit of the model may not be known in advance.

We also tested running a Gaussian process emulator using as input the single-parameter grid of simulations
performed for the quadratic emulator. This reveals whether the improvement of the Gaussian process over the quadratic emulator was due to the sampling strategy or the more flexible interpolation algorithm. Since the kernel chosen for the Gaussian process includes a dot kernel, which is equivalent to linear interpolation, we expect the Gaussian process to do at least as well as quadratic emulation. Of our $6$ test simulations, $4$ test simulations achieved $\sim 1\%$ accuracy, similar to the full Gaussian process emulator.  However, the other $2$ test simulations showed $\sim 4\%$ accuracy, similar to the quadratic emulator, although the simulations with the poorest estimation were not the same.

With a single parameter grid the Gaussian process over-estimated the emulator error by a factor of $4-10$. This has a simple explanation. The quadratic emulator relies on varying each parameter in the emulator separately and combining them, which is reasonably accurate in practice. However, without simulations which vary multiple parameters at once
the Gaussian process does not know this and is thus unable to properly constrain the emulator error. It thus seems
that the majority of the improved emulator performance with our Gaussian process comes from the improved sampling strategy. We infer that the chief limitation of the quadratic polynomial interpolation is that it forces a sampling strategy with
a relatively limited coverage of parameter space, especially away from the central mean flux value.

\subsection{Posterior parameter constraints}
\label{sec:likeresults}

\begin{figure}
\includegraphics[width=0.95\textwidth]{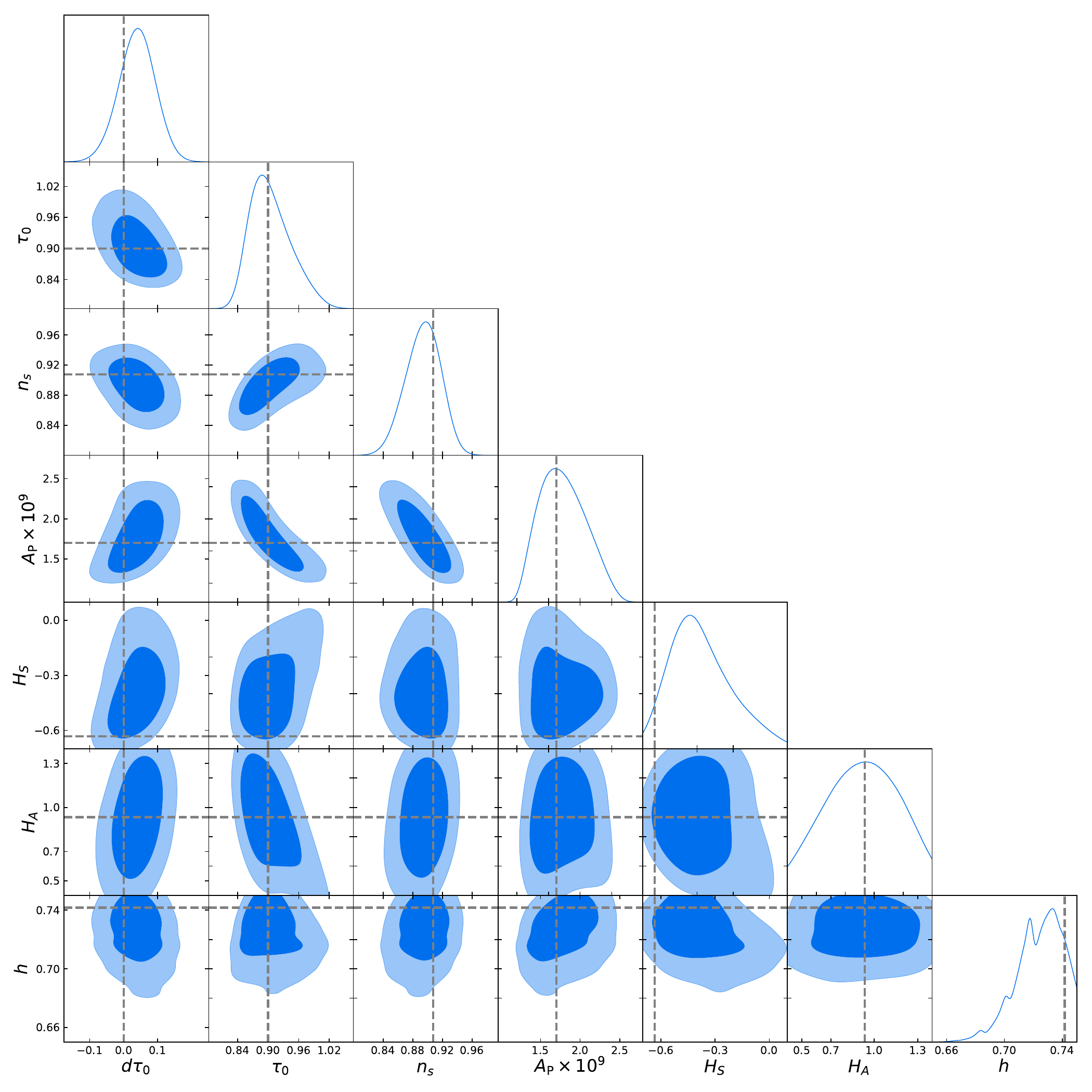}
  \caption{Marginalized likelihood posteriors in $1D$ and $2D$ for the parameters of our test simulation with $\tau_0 = 0.9$, $d\tau_0 = 0$, $n_s = 0.907$, $A_s=1.7\times 10^{-9}$,
  $H_A = 0.933$, $H_S = -0.633$, $h = 0.742$ using a $21$ sample Gaussian process emulator. Contours show $68\%$ and $95\%$ confidence intervals.
  The true parameter values are shown by dashed lines.}
  \label{fig:gplikelihood}
\end{figure}

\begin{figure}
\includegraphics[width=0.95\textwidth]{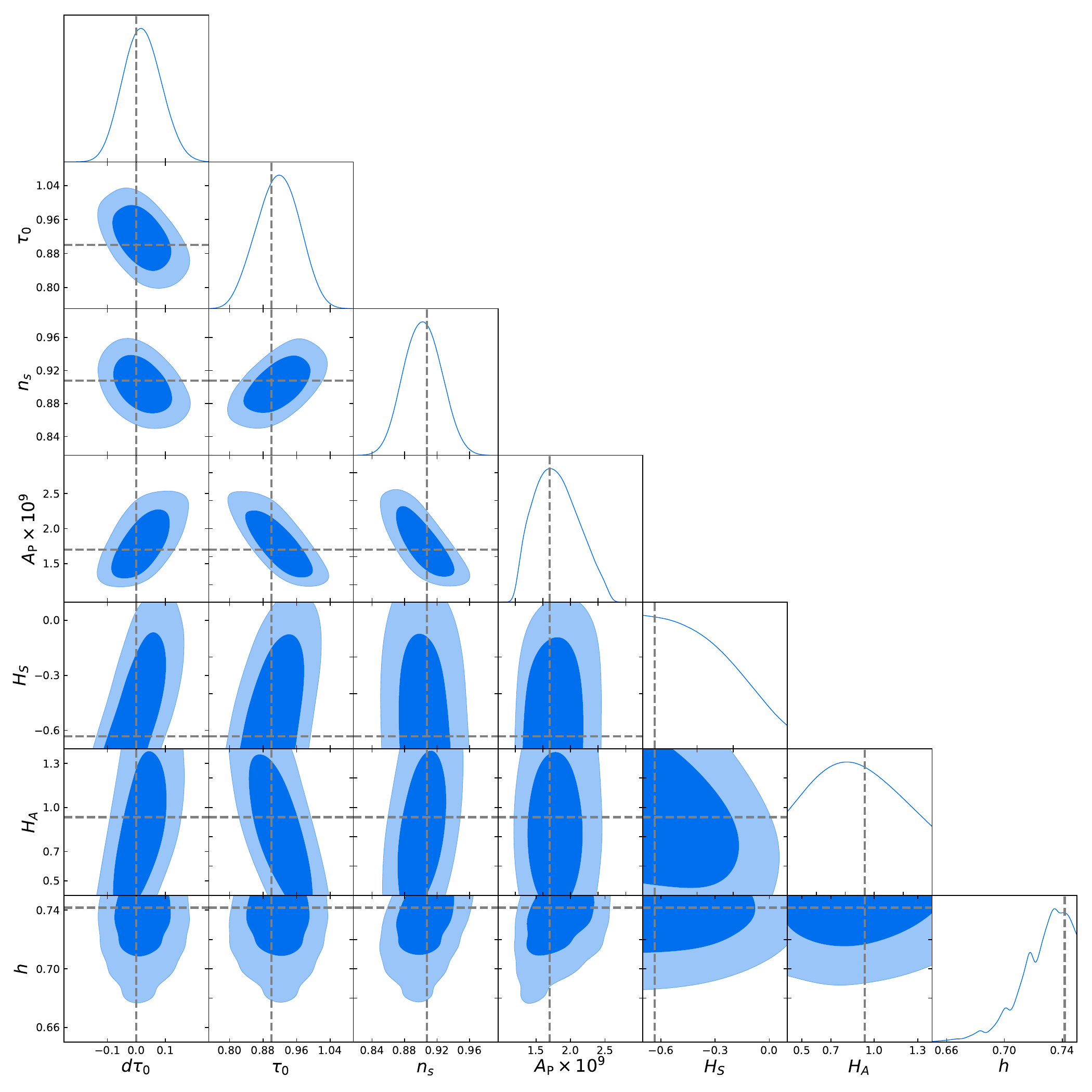} \\
  \caption{Marginalized likelihood posteriors in $1D$ and $2D$ for the parameters of our test simulation with $\tau_0 = 0.9$, $d\tau_0 = 0$, $n_s = 0.907$, $A_s=1.7\times 10^{-9}$,
  $H_A = 0.933$, $H_S = -0.633$, $h = 0.742$ using a $21$ sample quadratic polynomial emulator. Contours show $68\%$ and $95\%$ confidence intervals.
  The true parameter values are shown by dashed lines.}
  \label{fig:quadlikelihood}
\end{figure}


\begin{table}
\begin{center}
\begin{tabular}{|c|c|c|c|}
\hline
Parameter & Gaussian process & Quadratic & Quad/GP (\%)\\
\hline
$d \tau_0 $ & $-0.014$ - $0.092$ & $-0.046$ - $0.086$ & $25$\\
$\tau_0 $ & $0.866$ - $0.947$ & $0.864$ - $0.967$ & $26$\\
$n_s$ & $0.87$ - $0.92$ & $0.88$ - $0.97$ & $-3$ \\
$A_\mathrm{P} \times 10^9$ & $1.484$ - $2.094$ & $1.464$ - $2.140$ & $11$\\
$H_S$ & $-0.54$ - $-0.20$ & $-0.61$ - $-0.15$ & $34$ \\
$H_A$ & $0.65$ - $1.18$ & $0.58$ - $1.19$ & $16$ \\
$h$ & $0.71$ - $0.74$ & $0.71$ - $0.74$ & $4$ \\
\hline
\end{tabular}
\end{center}
\caption{Table of $68\%$ posterior parameter confidence intervals from the Gaussian process and quadratic emulators. The final column shows the percentage change in the confidence interval widths. Positive numbers indicate wider intervals in the quadratic emulator. The true parameter values of the input simulation are $\tau_0 = 0.9$, $d\tau_0 = 0$, $n_s = 0.907$, $A_s=1.7\times 10^{-9}$, $H_A = 0.933$, $H_S = -0.633$, $h = 0.742$ }
\label{tab:posteriors}
\end{table}

\begin{figure}
  \includegraphics[width=0.95\textwidth]{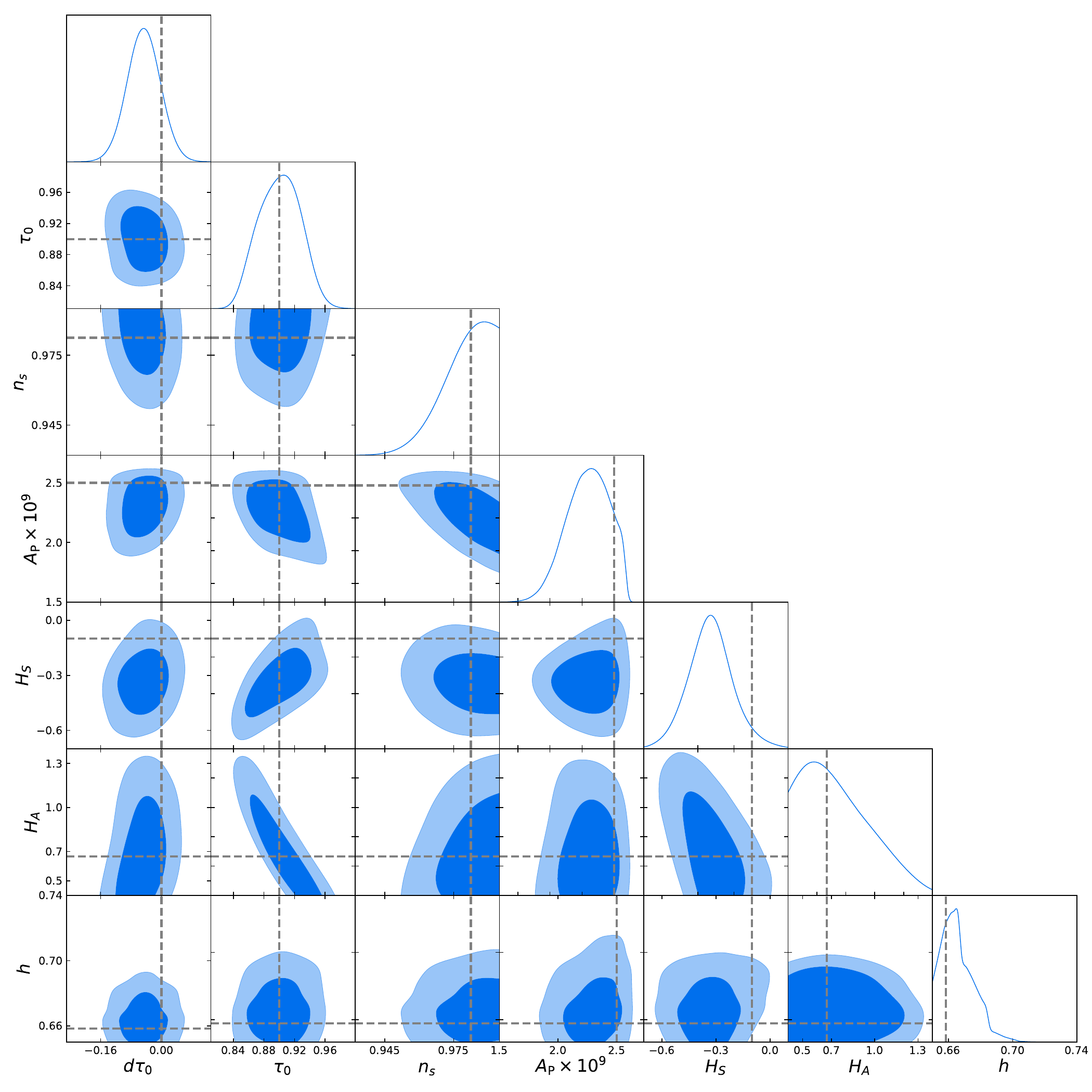}
  \caption{Marginalised likelihood posteriors in $1D$ and $2D$ for the parameters of our test simulation with $\tau_0 = 0.9$, $d\tau_0 = 0$,  $n_s = 0.982$, $A_s=2.5 \times 10^{-9}$, $H_A = 0.667$, $H_S = -0.1$, $h = 0.658$ using a $21$ sample Gaussian process emulator. Contours show $68\%$ and $95\%$ confidence intervals.
  The true parameter values are shown by dashed lines.}
  \label{fig:worstlikelihood}
\end{figure}



Figure~\ref{fig:gplikelihood} shows the $1D$ and $2D$ posterior parameter constraints from the Gaussian process emulator on a test simulation.
The emulator error at the true data points of the simulation is shown in figure~\ref{fig:nss8acc}, and is typical for test simulations using this emulator.
With emulator errors at this level ($1-2\%$ of the flux power spectrum), the posterior confidence intervals provide an unbiased measurement of the
true parameter values. Some of the thermal parameters are poorly constrained by the BOSS data. In particular the $95\%$ marginalized confidence intervals on $H_A$ and $H_S$
overlap with the emulator prior volume. Ref.~\cite{PD2015} find $\Delta\gamma = 0.2$ and $\Delta T_0 = 3500$ K at $68\%$ confidence.
When translated into constraints on $\gamma$ and $T_0$, our prior volume has $\Delta\gamma = 0.4$ and $\Delta T_0 = 9000$ K (see section~\ref{sec:simparams}), so our results are consistent. These relatively poor constraints are expected with the BOSS DR9 covariance, which does not measure the small scales that are most sensitive to the thermal history.
We considered enlarging our prior volume, but there are strong physical reasons to expect $\gamma$ and $T_0$ to be within this range. Alternatively, we considered placing a prior
on $T_0$ from high resolution flux power spectrum measurements, but again decided against it because we wished to demonstrate that our emulator produces unbiased results throughout the prior volume. Our constraints on $\tau_0$ are much tighter than our prior volume.

The Gaussian process emulator shows parameter degeneracies between $A_\mathrm{P}$, $n_s$ and $\tau_0$, including a curved degeneracy between $A_\mathrm{P}$ and $\tau_0$.
This is to be expected, since $A_\mathrm{P}$ and $\tau_0$ both set the overall amplitude of the flux power spectrum. The degeneracy between $n_s$ and $A_\mathrm{P}$ suggests that,
once the thermal parameters are marginalized out, the scale at which the \Lya~forest is most sensitive to cosmology is larger than the pivot scale we chose, $k_\mathrm{P} = 0.78$ Mpc$^{-1}$. This motivates a smaller $k_\mathrm{P}$ in future work.

Our constraints on $h$ appear unusually tight: $\Delta h = 0.04$ at $68\%$ confidence ignoring prior volume effects. They also exhibit unusual non-smooth features in the likelihood.
Recall from section~\ref{sec:simparams} that we hold $\Omega_M h^2$ constant and so a variation in $h$ is also a variation in $\Omega_M$. Most of the constraining power is thus coming from the effect of $\Omega_M$ on the growth rate of the cosmological perturbations at this redshift. The non-smooth likelihood features arise from our choice to ignore data on scales larger than our simulation box size. Since the mapping between the size of the simulation box (in comoving Mpc/h) to the measured bins of the flux power spectrum (in physical km/s) depends on $\Omega_M$ (eq.~\ref{eq:kms}), decreasing $h$ causes an extra flux power spectrum bin to be included in the likelihood, suddenly reducing the error bars. This feature would not occur in an emulator with a larger box.

Figure~\ref{fig:quadlikelihood} shows the posterior parameter constraints using the quadratic emulator. The true parameter values again all lie within the $68\%$ confidence intervals, but the constraints on key cosmological parameters are substantially weaker. The $68\%$ marginalized confidence intervals are given in Table~\ref{tab:posteriors}. Increases in the confidence interval widths are largest in the thermal parameters, $H_S$ and $H_A$, which have the most non-linear effect on the flux power spectrum (see Appendix~\ref{sec:singleparam}). Constraints on the primordial power spectrum amplitude, $A_\mathrm{P}$ are also weaker, due to the degeneracy between $A_\mathrm{P}$ and $\tau_0$.

The quadratic polynomial emulator has larger emulator error, but our likelihood function does not include it in the covariance matrix. We might thus naively expect a biased parameter value rather than larger uncertainty. However, the true change in the flux power spectrum at parameter values away from the central ``best-fit'' simulation for several parameters, in particular $\tau_0$, is larger than estimated by the quadratic polynomial. This leads to the emulator under-estimating the constraining power of the data, leading to larger uncertainties, and these larger uncertainties are sufficient to dominate over biased parameters from an inaccurate likelihood.
Note that the magnitude of the difference in posterior errors depends on the specific prior volume and dataset used.

Figure~\ref{fig:worstlikelihood} shows the posterior likelihood constraints for the second test simulation shown in figure~\ref{fig:nss8acc}, for a Gaussian process emulator. This simulation serves as a useful worst-case scenario for our Gaussian process emulator as this test simulation had the largest error between the emulated and true flux power spectra. In this case the emulator error is $3-4\%$, comparable to the diagonal elements of the BOSS covariance matrix at low redshift. However, with the exception of $H_S$, the true parameter values are within the $68\%$ confidence intervals, demonstrating that the emulator produces unbiased parameters. We examined the posterior contours for the other $4$ test simulations and in all other cases the true parameter values lay within the $68\%$ confidence intervals. Furthermore, in all cases the quadratic polynomial emulator produced weaker constraints than the Gaussian process emulator, with the increase in the size of the contour widths similar to those already discussed.

We checked the effect of artificially setting emulator error to zero in the likelihood function using our test simulations. For the worst-case test simulation shown in figure~\ref{fig:worstlikelihood} posterior confidence interval widths shifted by $\sim 10\%$, with the constraints on $n_s$ tightening and those on $H_S$ weakening.
The posterior contours continued to enclose the true parameter values within the $68\%$ confidence intervals. We caution that this continued unbiased estimation of the posterior may simply be because the emulator is moderately over-estimating the true error for this test point (see figure~\ref{fig:erracc}). For our other test simulations there was minimal effect on the
posterior confidence intervals, probably because the emulator error in these cases was $< 50\%$ of the statistical error from BOSS DR9.


\section{Conclusions}
\label{sec:conclusions}

We have constructed and validated techniques for emulation of the \Lya~forest 1D flux power spectrum. We have not yet performed simulations of sufficient
size to produce a resolved flux power spectrum, instead using this paper to validate our emulator building routines\footnote{Publicly available at \url{https://github.com/sbird/lya_emulator/}} with a series of small, $40$ Mpc/h, simulations with $2\times 256^3$ particles.
We sample $5$ parameters with $21$ simulations in a Latin hypercube scheme. There are $2$ parameters describing the thermal history of the IGM and $3$ describing cosmology. We add a further $2$ parameters for the mean flux in post-processing. Our emulator allows the generation of 1D \Lya~forest flux power spectra at arbitrary parameter values within
the emulator volume, unlike existing interpolation methods which only perform well near a fiducial simulation. A Gaussian process is used for interpolation, with a kernel including both linear and radial basis function terms.

We tested our emulator with a series of $6$ simulations spread throughout parameter space. The typical interpolation error was $1\%$, with a worst-case performance of $4\%$, comparable to the BOSS DR9 1D \Lya~forest flux power spectrum statistical error of $3-10\%$ \cite{PD2013}. The Gaussian process estimates the interpolation error as well as the function value at the interpolation point. We demonstrated that in our test simulations the estimate of this error is reasonably reliable. It was accurate in $4$ of our $6$ tests and moderately larger than the true emulator error in the other $2$.

We constructed a likelihood function for our emulator using our test simulations and the BOSS DR9 data covariance matrix as mock data. The likelihood function propagated the estimated emulator error from the Gaussian process, and thus accounted for residual interpolation error. We showed that MCMC chains run using our likelihood function were able to recover the true parameters of the test simulation, within $68\%$ confidence intervals. Interestingly, an earlier test emulator with a prior volume five times larger did not produce unbiased posterior results, and showed distinctly non-Gaussian error residuals. This suggests that for the Gaussian process to produce an unbiased posterior estimate requires a critical density of simulation points. We suggest that the this density must be sufficient for accurate optimization of kernel hyper-parameters. We considered the effect of artificially setting the expected emulator error to zero in our likelihood, and discovered that it was small in most cases, as for our emulator the estimated error is much smaller than the data covariance.

We compared our Gaussian process emulator to a quadratic polynomial emulator resembling those used by earlier \Lya~forest analyses, showing that it produced substantially smaller emulator error (by a factor of $2$) with the same number of simulations and prior volume. To evaluate the impact on posterior parameter constraints, we ran MCMC chains using our quadratic emulator and showed that posterior confidence intervals using a typical test simulation were on average $16\%$ wider. We (reassuringly) showed that the quadratic emulator did not produce biased parameter constraints. The emulator error for quadratic interpolation was around $4\%$, which suggests that although it is sufficient for data with statistical uncertainties of $\sim 10\%$ such as in ref.~\cite{Irsic:2017c}, it will be inadequate for future datasets with smaller statistical error. The techniques and code we present in this work will thus be essential for making best use of future \Lya~forest surveys.

Similar emulators would also be useful for Lyman-$\alpha$ forest datasets derived from high resolution spectra. These datasets probe smaller scales, up to $k = 0.2$ s/km \citep{Boera:2018} and are generally more sensitive to the details of reionization and the thermal history of the IGM. The simulations presented here are not of sufficient resolution to probe these scales with fidelity, so we have not attempted to build an emulator for them here. However, it should be possible to build a similar emulator to ours for these scales using higher resolution simulations and a model for reionization.

Although our focus in this paper has been the \Lya~forest, the techniques of cosmological emulation are generally applicable to any problem which requires expensive forward simulation, in particular those sensitive to non-linear gravitational evolution. Such problems include galaxy clustering, weak lensing, 21cm and the halo mass function, most of which have had emulators constructed for them. The most generally applicable points in this paper are our use of a flexible Latin hypercube design which maximizes the coverage of simulation points, our understanding of the importance of good hyper-parameter estimation and that this can be improved using multi-output Gaussian processes.

In a companion paper, ref.~\cite{Rogers:2019}, we show how the \Lya~emulator can be improved by Bayesian emulator optimization, which iteratively adds points to the simulation sample set, and produces further improvements in the width of posterior confidence intervals. In these two papers we have demonstrated that the techniques presented here are capable of achieving the percent-level accuracy necessary for an optimal analysis of future \Lya~forest surveys, such as DESI. In future work we will apply our techniques to building a full cosmological emulator for the \Lya~forest. We will also incorporate more sophisticated models of hydrogen and helium reionization, which will allow us to improve the fidelity of our modelling as well as the interpolation error.

\acknowledgments

SB was supported by NSF grant AST-1817256. Computing resources were provided by NSF XSEDE allocation AST180021 and the Maryland Advanced Research Computing Center (MARCC). KKR was supported by the Science Research Council (VR) of Sweden. HVP was partially supported by the European Research Council (ERC) under the European Community's Seventh Framework Programme (FP7/2007-2013)/ERC grant agreement number 306478-CosmicDawn, and the research project grant ``Fundamental Physics from Cosmological Surveys'' funded by the Swedish Research Council (VR) under Dnr 2017-04212. AP was supported by the Royal Society. HVP and AP were also partially supported by a grant from the Simons Foundation. LV was supported by the European Union's Horizon 2020 research and innovation programme ERC (BePreSySe, grant agreement 725327) and Spanish MINECO projects AYA2014-58747-P AEI/FEDER, UE, and MDM-2014-0369 of ICCUB (Unidad de Excelencia Maria de Maeztu). AFR was supported by a Science and Technology Facilities Council (STFC) Ernest Rutherford Fellowship, grant reference ST/N003853/1. HVP, AP and AFR were further supported by STFC Consolidated Grant number ST/R000476/1. This work was partially enabled by funding from the University College London (UCL) Cosmoparticle Initiative.

\appendix
\section{Single-parameter variations}
\label{sec:singleparam}

\begin{figure}
  \includegraphics[width=0.45\textwidth]{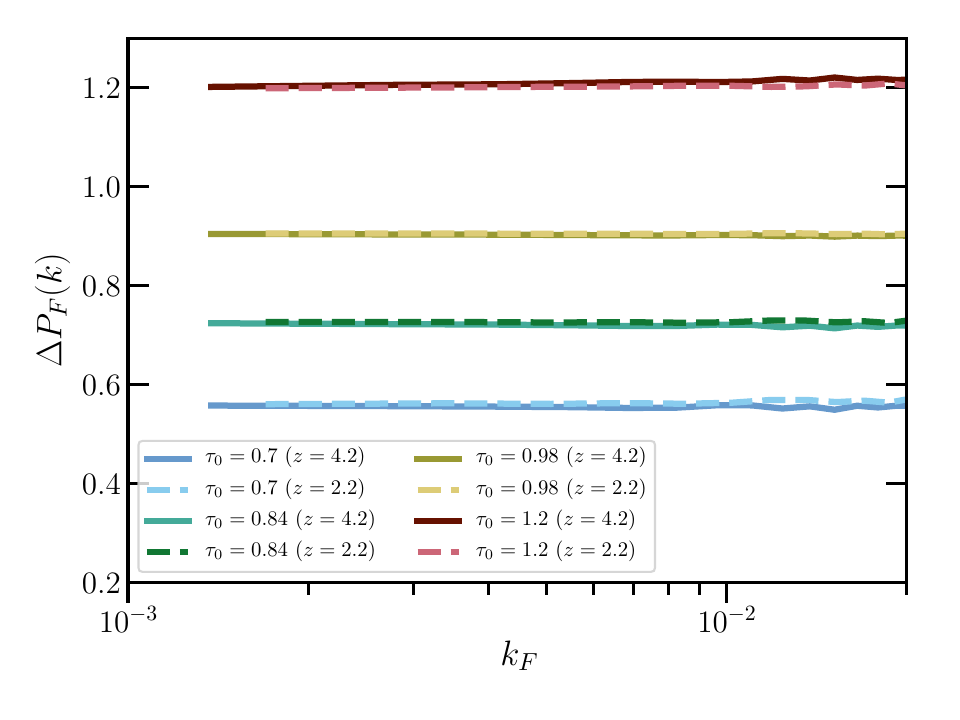}
    \includegraphics[width=0.45\textwidth]{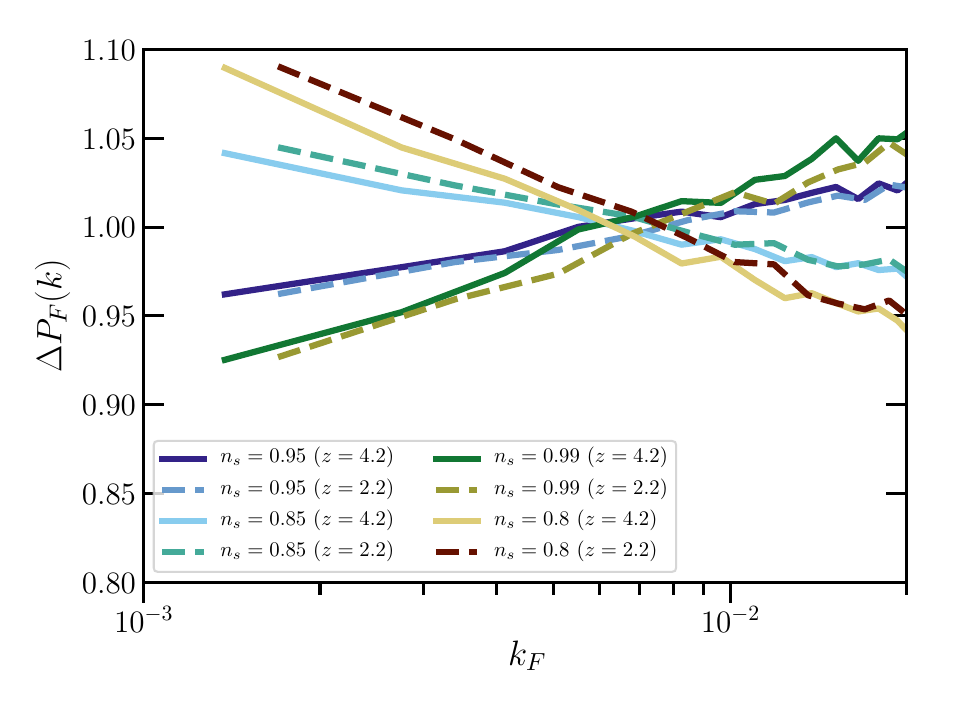} \\
  \includegraphics[width=0.45\textwidth]{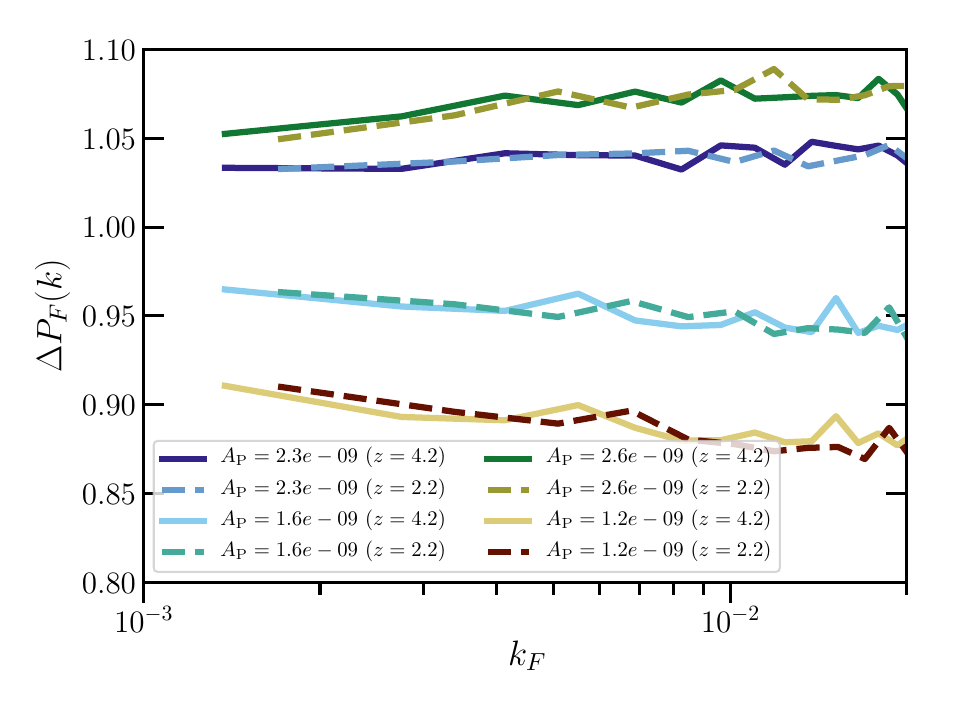}
  \includegraphics[width=0.45\textwidth]{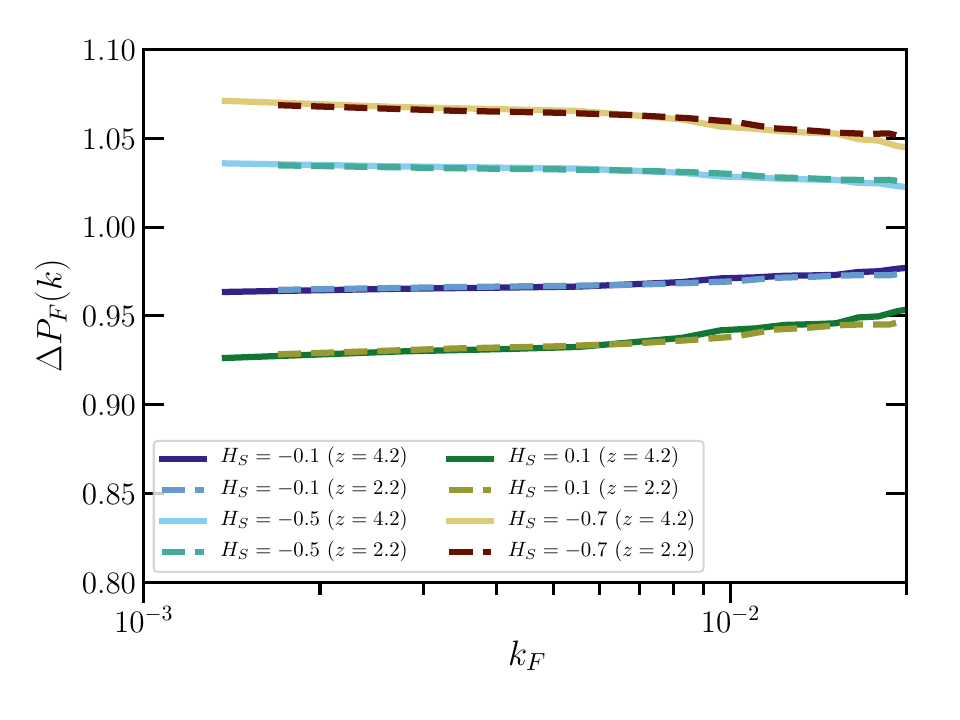} \\
  \includegraphics[width=0.45\textwidth]{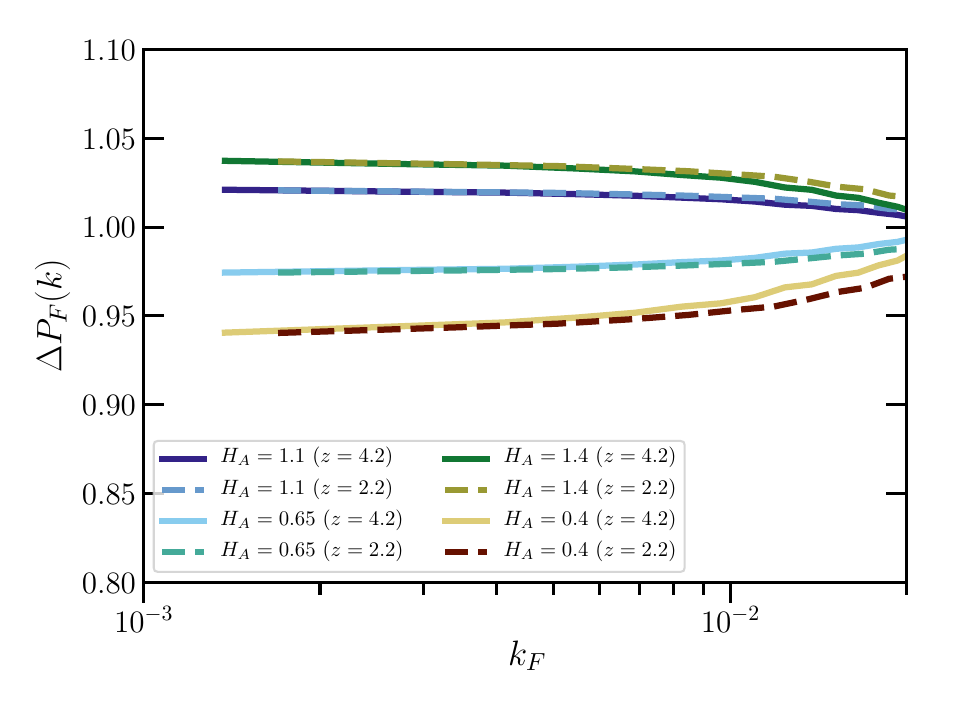}
  \includegraphics[width=0.45\textwidth]{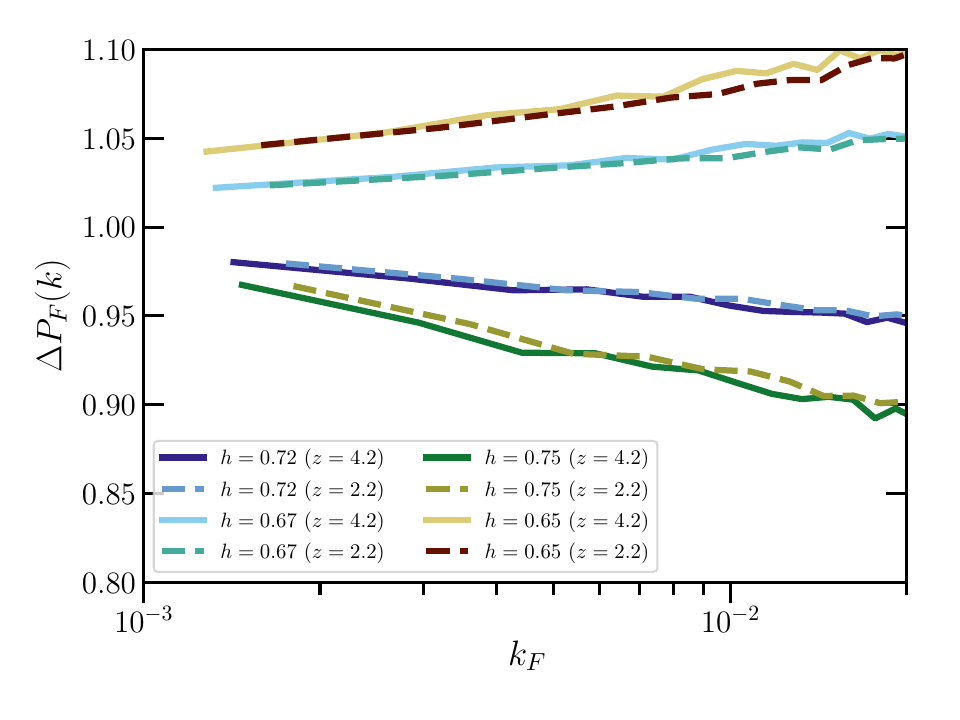}
  \caption{The effects of changing each emulator parameter on the 1D flux power spectrum. Shown is the ratio between the flux power spectrum at each parameter value used for the quadratic emulator to the ``best-fit'' simulation at $\tau_0 = 1$, $d\tau_0 = 0$, $n_s = 0.8975$, $A_\mathrm{P} = 1.9 \times 10^{-9}$, $H_S = -0.3$, $H_A = 0.9$, $h = 0.7$. The parameters are: (top-left) the mean flux $\tau_0$, (top-right) the spectral index $n_s$, (middle-left) the primordial power spectrum amplitude $A_\mathrm{P}$, (middle-right) the heat slope $H_S$, which controls the temperature-density relation parameter $\gamma$, (bottom-left) the heat amplitude $H_A$, which controls the temperature at mean density $T_0$ and (bottom-right) the hubble parameter, $h$, which is inversely proportional to the matter density $\Omega_M$ in our scheme. We show the minimum and maximum redshifts in the emulator: $z=2.2$ (dashed) and $z=4.2$ (solid).}
  \label{fig:Aschange}
\end{figure}

Figure \ref{fig:Aschange} shows the effect on the 1D flux power spectrum of changing a single parameter
in our emulator, using the simulations from the quadratic polynomial emulator.
While this has been discussed in the literature before \citep[e.g.~ref~][]{Viel:2006}, we have interpolated a
non-standard set of parameters, and so in the interests of clarity we show how our new parameters affect the flux power spectrum.
We show the effect at $z=2.2$ and $z=4.2$ to demonstrate that the change in each parameter with redshift is relatively small.
The redshift-dependent conversion between km/s and comoving Mpc/h is visible as
specific features, including the box size, move with redshift. The degeneracy between $\tau_0$ and $A_\mathrm{P}$ is also visible,
although this would be partially broken by the largest scale bin if our box size was increased. We have rescaled each spectrum to have the same mean flux. Increasing the IGM temperature using $H_A$ increases the flux power spectrum only on the relatively large scales shown (and only because we must match the mean flux). On smaller scales it erases power, as expected.

\bibliographystyle{JHEP}
\bibliography{surrogate_refs}
\end{document}